\documentclass[a4paper]{article}
\usepackage[a4paper, total={6.2in, 9in}]{geometry}
\usepackage{floatrow}
\usepackage{hyperref}
\usepackage{filecontents}
\usepackage{subfigure}
\usepackage{bm}
\usepackage[font={scriptsize,bf}]{caption}
\usepackage{amsmath}
\usepackage{amsfonts}
\usepackage{amssymb}
\usepackage{graphicx}
\usepackage{bm}
\usepackage{multicol}
\usepackage{multirow}
\usepackage{epstopdf}
\usepackage{soul}
\usepackage{pifont}
\usepackage{lineno}
\usepackage[table]{xcolor}
\usepackage{color}
\usepackage[autostyle]{csquotes}
\usepackage{tikz,pgf}
\usepackage{collcell}

\usepackage{pgfplots}
\usepackage{wasysym}
\usepackage{lipsum}
\usepackage{algorithm} 
\usepackage{algpseudocode} 
\newlength\fheight
\newlength\fwidth
\newcommand{\cmmnt}[1]{\ignorespaces}
\usepackage{etoolbox}
\usepackage{setspace}
\usepackage{csquotes}

\title{Aeroelastic Reduced-Order Model Differential Equations in Transonic Buffeting Flow}
\author{ \large Michael Candon$^1$\thanks{corresponding author, candon.michael@rmit.edu.au}, Pier Marzocca$^1$ and Earl Dowell$^2$}

\date{
	\normalsize $^1$Department of Aerospace Engineering, RMIT University, Melbourne, AUS, 3000\\
	$^2$Duke Universirty, Durham, NC, 27708\\[2ex]%
}

\begin{document}
	
	\maketitle
	\begin{abstract}
        Numerical simulation of transonic shock buffet remains a formidable challenge due to its inherently nonlinear, unsteady nature. These difficulties are further compounded for three-dimensional configurations and when aeroelastic coupling is included. Consequently, computational studies of aeroelastic shock buffet interactions have largely been confined to two-dimensional systems. This limitation motivates reduced-order models (ROMs) that can efficiently and accurately reproduce aeroelastic responses in buffeting flow. This paper presents a nonlinear unsteady aerodynamic ROM that combines self-excited nonlinear oscillator dynamics with Volterra theory to represent nonlinear memory effects. The resulting integro–differential equation ROM (IDE-ROM) is identified using orthogonal matching pursuit (OMP). Application to the ONERA OAT15A airfoil demonstrates that the compact formulation reproduces key nonlinear behaviors, including aeroelastic lock-in, with high accuracy relative to CFD/CSD benchmarks. It is shown that lock-in occurs when buffet-induced aerodynamic damping overcomes structural damping, yielding negative net damping. Consistent with this interpretation, aerodynamic damping extracted from forced harmonic excitation is shown to recover the lock-in region and LCO amplitude without repeated aeroelastic simulations. The limitations and potential extensions of the approach are also discussed.
        \end{abstract}

        \section*{Nomenclature}
        \begin{tabbing}
        XXXXXXXXXX \= XX \= \kill
        $b$ \>\> Semi-chord [m]\\
        $c$ \>\> Chord [m]\\
        $c_{a_h}$, $c_{a_\alpha}$ \>\> Aerodynamic damping in heave [N$\cdot$s/m] and pitch [N$\cdot$m$\cdot$s/$^\circ$]\\
        $\bm{c}$ \>\> Vector containing identified ODE coefficients\\
        $C_L$, $C_M$, $C_p$  \>\> Lift, moment and pressure coefficient\\
        $\Delta C_L$ \>\> Lift coefficient peak-to-peak amplitude\\
        
        $\hat{f}$ \>\> Frequency ratio, $\omega/\omega_{B}$\\
        $h$, $\dot{h}$, $\ddot{h}$ \>\> Heave displacement, velocity and acceleration [m],  [m/s],  [m/s$^2$]\\
        $\hat{h}$ \>\> Amplitude of forced harmonic excitation in heave [m]\\
        $H_1$ \>\> Frequency response estimator \\
        $\mathcal{H}_j$ \>\> Vector containing $j^{th}$-order pruned Volterra series coefficients\\
        $I_\alpha$ \>\> Airfoil sectional moment of inertia [kg$\cdot$m]\\
        $k_{a_h}, k_{a_\alpha}$ \>\> Aerodynamic stiffness in heave [N/m] and pitch [N$\cdot$m/$^\circ$] \\
        $K$ \>\> Effective coupling gain in Adler equation\\
        $\bm{L}$ \>\> Lower left triangular circulant matrix\\
        $m$ \>\> Airfoil mass per unit span [kg/m]\\
        $M_\infty$ \>\> Freestream Mach number\\
        $N_O$, $N_I$ \>\> Number of candidate differential and integro-differential equation terms\\
        $N_L$ \>\> Number of time lags\\
        $N$ \>\> Total number of training samples\\

        $p$ \>\> Pruned Volterra series order\\
        $p_O$ \>\> Polynomial expansion order\\
        $q_\infty$ \>\> Dynamic pressure, $q_\infty=\tfrac{1}{2}\rho_\infty v_\infty^2$ [Pa]\\
        $Q$, $\dot{Q}$, $\ddot{Q}$ \>\> Generalized aerodynamic force, velocity and acceleration [-],  [1/s],  [1/s$^2$]\\
        $Re_\infty$ \>\> Freestream Reynolds number \\
        $t$ \>\> Time [s]\\
        $u$, $\dot{u}$, $\ddot{u}$ \>\> Generalized displacement, velocity and acceleration [-],  [1/s],  [1/s$^2$]\\
        $u_\infty$ \>\> Freestream velocity [m/s]\\
        $W$ \>\> Aerodynamic work per cycle\\
        $y^+$ \>\> Non-dimensional first cell height\\
        \vspace{1cm}\\
        \textbf{Greek Symbols}\\
        $\alpha$, $\dot{\alpha}$, $\ddot{\alpha}$ \>\> Pitch rotation, rotational velocity and rotational acceleration [$^\circ$],  [$^\circ$/s],  [$^\circ$/s$^2$]\\
        $\hat{\alpha}$ \>\> Amplitude of forced harmonic excitation in pitch [$^\circ$]\\
        $\alpha_0$ \>\> Freestream angle-of-attack [$^\circ$]\\

        $\epsilon$ \>\> Nonlinearity parameter in the Rayleigh oscillator\\
        $\kappa$ \>\> Pre-defined number of non-zero coefficients to be identified by OMP \\
        $\zeta_{h}$, $\zeta_{\alpha}$ \>\> Heave and pitch structural damping ratio\\
        $\mu$ \>\> Structural-to-fluid mass ratio, $m/\pi\rho_\infty b^2$\\
        $\rho_\infty$ \>\> Freestream fluid density [kg/m$^3$]\\
        $\tau$, $\Delta \tau$  \>\> Non-dimensional time and time-step $tu_\infty/c$, $\Delta tu_\infty/c$\\
        $\bm{\Phi}$, $\bm{\Phi_{NL}}$ \>\> Linear and nonlinear state matrix\\
        
        $\phi$ \>\> Phase difference between buffet and structural oscillations\\
        
        $\omega_{h}$, $\omega_{\alpha}$ \>\> Heave and pitch natural frequency [rad/s]\\
        $\omega_{B}$ \>\> Shock buffet frequency [rad/s]\\
        $\Delta \omega$ \>\> Frequency detuning ($\omega_B - \omega_h$ or $\omega_B - \omega_\alpha$) \\
        \end{tabbing}
        
        \section{Introduction}
        Transonic shock buffet is a nonlinear unsteady aerodynamic phenomenon, characterized by large-amplitude self-sustained periodic shock oscillations that result from shock-boundary layer interactions~\cite{giannelis17}. Shock buffet is a global flow instability, meaning that it occurs even in the absence of structural motion (referred to as a fluid-only limit cycle oscillation (LCO) throughout this paper). Although shock buffet occurs only within a narrow window of the transonic regime, the aeroelastic instabilities it induces make it a primary contributor to fatigue life degradation, while also limiting the flight envelope and adversely affecting pilot handling qualities and comfort~\cite{levinski20}. While a major driver for transonic buffet research comes from the defense sector ($e.g.$, the F-16 is renowned for its issues with the phenomenon~\cite{ionovich17}), the civil aviation sector also requires careful consideration of its effects. For instance, the sustainable aviation sector is seeing a demand for novel lightweight and aerodynamically efficient aircraft designs, requiring that practitioners pay very close attention to dynamic aeroelastic effects in the transonic regime - as exemplified in Boeing's work on the Transonic Truss-Braced Wing~\cite{browne25}. 
        
        The aerodynamic aspects of shock buffet on a rigid body have been widely studied, with experimental campaigns dating back to the 1980s~\cite{mcdevitt85}, and more recent campaigns considering the two-dimensional~\cite{jacquin09} and three-dimensional~\cite{koike16} buffet mechanism. In the last two decades, exponential growth in computing power and broader access to high-performance computing (HPC) have enabled researchers to study the phenomenon numerically, using computational fluid dynamics (CFD) codes. Such research was initially dominated by two-dimensional Unsteady Reynolds-averaged Navier-Stokes (URANS) simulations with mixed success. On one hand, researchers have demonstrated that URANS codes can capture two-dimensional buffet with good accuracy while, on the other hand, generalization largely remains problematic. In particular, different turbulence models can produce vastly different results. Most recently, numerical modeling of three-dimensional buffet has seen major interest on wings or half-span aircraft models. In such cases, it seems that the limitations of URANS codes are more prevalent~\cite{giannelis23}. As a result, scale-resolving codes, including variants of Large Eddy Simulation (LES), have seen increasing use for buffeting flows, albeit at a prohibitive computational cost. Understanding the interaction between shock buffet and an elastic structural model is also a critical aspect of this field of research that has made rapid progression in the last 15 years. Before examining this body of literature, the frequency ratio, $\hat{f}_n$, is defined as: 
        
        \begin{equation}
            \label{eq:1}
            \hat{f}_n = \frac{\omega_n}{\omega_{B}}
        \end{equation}
        
        \noindent where $\omega_n$ could be the structural natural frequency or forced oscillation frequency of mode $n$, and $\omega_{B}$ is the shock buffet frequency. One of the earliest studies of Raveh~\cite{raveh09} demonstrates that frequency lock-in can occur when an airfoil in transonic buffeting flow undergoes forced sinusoidal motion at frequency ratios in the vicinity of one, $i.e.$, the unsteady shock oscillations synchronize with the airfoil's structural motion. This lock-in to forced harmonic motion was confirmed experimentally by Hartmann~\cite{hartmann13} who also investigated the two-degree-of-freedom (2-DOF) heave-pitch aeroelastic response. The heave and pitch frequency ratios investigated by Hartmann were much less than one ($\hat{f}_h << 1$, $\hat{f}_\alpha << 1$), and lock-in was not observed, but rather the system was shown to respond at the buffet frequency. Subsequent numerical studies of two-dimensional elastically suspended airfoils in transonic buffet~\cite{raveh14, quan15, giannelis16} have shown that {\color{black}the unsteady aerodynamic loads can exhibit frequency lock-in to a single-degree-of-freedom (s-DOF) pitching mode when the structural frequency ratio is slightly above unity ($\hat{f}_\alpha \gtrsim 1$). In this regime, the shock oscillation synchronizes with the structural motion and the response can be strongly amplified near the structural natural frequency. Lock-off is typically observed as the frequency ratio increases further (often before $\hat{f}_\alpha \approx 2$), after which the response returns to a lower-amplitude oscillation dominated by the buffet frequency. Outside the lock-in region, structural oscillations are generally smaller and remain primarily driven at $\omega_B$.} These studies also assess sensitivities of lock-in to structural damping and to the structural-to-fluid mass ratio. Gao~\textit{et al.}~\cite{gao17} proposed a carefully designed linearized reduced-order model (ROM) to study lock-in, showing that it is in fact driven by a coupled mode flutter, rather than a form of resonance as previously thought. The coupling occurs between the unstable fluid mode and unstable structural mode. Gao and Zhang~\cite{gao20} go on to formally define different forms of transonic aeroelastic instabilities in terms of the interaction between structural and fluid modes. Limited work has been conducted which considers three-dimensional aeroelastic modeling~\cite{ionovich17}. Very recently the aeroelastic lock-in phenomenon has also been investigated experimentally~\cite{themiot23, korthauer23}.  
        
        When considering two-way coupled aeroelastic simulations in transonic buffeting flow the aforementioned computational overhead is increased substantially. Phenomena of interest, such as lock-in, can take much time to develop, requiring coupled CFD/CSD simulations of hundreds of thousands or even millions of time steps. Extension to three-dimensional buffet aeroelastic problems can quickly become computationally intractable. If supercomputing resources are available it becomes more feasible, however, supercomputing is expensive and not all investigators have access to such facilities. If computationally efficient reduced-order methods were available for this class of problems, it would allow for rigorous studies of the influence of different aeroelastic parameters, and of aeroelastic systems of greater complexity. 

        Nonlinear ROMs for unsteady aerodynamic and aeroelastic systems have progressed significantly in the last half-century, including; Volterra theory~\cite{silva05}, nonlinear oscillator models~\cite{hartlen70}, proper orthogonal decomposition~\cite{hall00}, dynamic mode decomposition~\cite{fonzi24}, harmonic balance~\cite{thomas04a}, and the broad class of projection-based methods~\cite{carlberg13}. ROM approaches for transonic buffet are a relatively new proposition that have not yet seen widespread attention, particularly within the realm of aeroelasticity. Although the work of Gao \textit{et al.}~\cite{gao17} that was described previously uses a buffet aeroelastic ROM, it is linear and only intended to assess stability (not to capture the aeroelastic LCO). In terms of a nonlinear unsteady aerodynamic ROM that can be used to model the full aeroelastic response of a system in buffeting flow, the challenges are significant. Critical components of such a ROM include the ability to capture:
        
        \begin{enumerate}
            \item Self-excited and sustained fluid instabilities (fluid LCO) in the absence of structural motion.
            \item Fluid-structure coupling with phenomena like flutter, LCO and lock-in.  
            \item Nonlinear memory effects in the generalized forces due to large scale transonic shock dynamics and separation. 
        \end{enumerate}
        
        One important consideration is that only the airfoil or wing surface flow quantities are of real interest - simplifying the problem to some degree. Approaches based on integrated quantities (forces and moments) are attractive as it keeps training dimensionality low. Two approaches will now be interrogated further: those based on Volterra theory and nonlinear oscillator models. When it comes to self-excited flow-only LCOs, an analogy that has been studied for decades should be considered; the aeroelastic response to vortex shedding over a bluff body. Several authors have shown that the unsteady aerodynamic forces in this scenario can be phenomenologically described by canonical nonlinear oscillator models~\cite{hartlen70,skop73,dowell81,marra11,hollenbach21}. Under such a formulation, the structural dynamics is described by the standard structural equation of motion, $i.e.$, an oscillator with mass, stiffness, and damping, and the fluid dynamics is described by a separate nonlinear fluid oscillator model. The fluid oscillator can take the form of a nonlinear second-order ordinary differential equation (ODE). A relatively straightforward example of this class of ROM is described by Dowell~\cite{dowell81}, where it is demonstrated that by combining a Van der Pol oscillator with the Parkinson galloping model, the transverse structural dynamic response of a bluff body encountering vortex shedding can be described. The Van der Pol oscillator will be discussed in greater detail later in the paper. While being entirely relevant and suited to buffet, such a model may not, on its own, account for the pronounced nonlinear memory effects that arise due to large-amplitude transonic shock motion (the third item described above). One approach that is well suited to capture transonic aerodynamic nonlinearities in the realm of aeroelasticity is the Volterra series~\cite{silva97}, expressing the aerodynamic forces as a functional series of convolutions of the structural motion with multi-dimensional kernels that capture memory and nonlinear behavior. Recent efforts in data-driven identification of sparse Volterra kernels have expanded its use case to systems with more intense nonlinear aerodynamics and higher dimensionality~\cite{balajewicz10, depaula19, candon24b, candon24c, candon25a}. However, on its own, the Volterra series is not useful as it cannot capture self-excitation of the fluid. 
        
        A combination of these two ROM paradigms seems logical. One could envisage the formulation of an integro-differential equation (IDE) where a nonlinear oscillator model (ODE) handles flow-only LCO and the Volterra series (integral equation) is embedded to improve the ability of the model to capture higher-order nonlinear memory effects. The question is how to identify such a model. A contemporary approach for the identification of dynamical systems from data was introduced by Brunton~\textit{et al.}, who proposed the Sparse Identification of Nonlinear Dynamics (SINDy)~\cite{brunton16} framework for the discovery of compact interpretable equations from data. SINDy has had a profound impact within the dynamical systems research communities, with relevant applications including the identification of the canonical flow past cylinder~\cite{loiseau18} problem, and more recently to identify nonlinear ODEs to describe the unsteady forces associated with shock buffet-only (not considering the aeroelastic response)~\cite{sansica22, ma25}. To the authors' knowledge, nonlinear unsteady aerodynamic ROMs for shock buffet when applied to an aeroelastic system have not been published in the open literature. 
        
        This paper proposes a simple yet effective framework for the identification of nonlinear ROMs from data, for aeroelastic shock buffet interactions. The proposed ROM combines the traditional concepts of nonlinear oscillator models to describe self-excited fluid flows, and Volterra series models to describe nonlinear transonic flow phenomena, with contemporary system identification / machine learning approaches to discover the system of equations from data. The SINDy algorithm is not used per se, but instead sparsity is promoted via greedy selection using Orthogonal Matching Pursuit (OMP). The objective is to identify a single set of reduced order model nonlinear differential equations that can describe the unsteady forces and moments due to buffeting flow on stationary, oscillating, and elastic airfoil models.

\section{Reduced-Order Model Differential Equations}
\label{sec:DEROM}

For consistency with the CFD datasets and the identification procedure, all models in this section are written in discrete time. Time histories are sampled at uniform intervals $t_n=n\Delta t$ and $x_n := x(t_n)$. The quantities $\dot{x}_n$ and $\ddot{x}_n$ represent numerical estimates of the first and second time derivatives at $t_n$ (computed from the sampled data using finite differences).

\subsection{Model structure}

Let $Q_n$ denote a generalized aerodynamic force and let $u_n$ denote a generalized structural coordinate. In its simplest form, the aerodynamic-force acceleration at time step $n$ is given as a function of (i) the current aerodynamic state $(Q_n,\dot{Q}_n)$, and (ii) the current and past structural motion. This can be written as the sum of a fluid-oscillator component $\mathcal{F}_F(\cdot)$ that depends only on $(Q_n,\dot{Q}_n)$ and a structural motion-induced component $\mathcal{F}_S(\cdot)$ that depends on a structural history vector: 

\begin{equation}
\ddot{Q}_n =
\mathcal{F}_F\!\left(Q_n,\dot{Q}_n\right)
+
\mathcal{F}_S\!\left(\bm{z}_n\right)
\label{eq:4}
\end{equation}

\noindent where $\bm{z}_n = \left\{u_{n-k},\dot{u}_{n-k},\ddot{u}_{n-k}\right\}_{k=0}^{N_L}$ is the structural history vector truncated for $N_L$ time lags. In the simplest case, $\mathcal{F}_F(\cdot)$ can be chosen as a canonical nonlinear oscillator ($e.g.$, Rayleigh/Van der Pol type) to permit a self-sustained buffet limit cycle, while $\mathcal{F}_S(\cdot)$ can be represented using Parkinson's galloping model~\cite{parkinson89} (current \textit{lag-0} structural states only, $N_L = 0$) or a pruned finite memory Volterra series~\cite{balajewicz12}. If necessary, $\mathcal{F}_F(\cdot)$ may also be discovered from a library of candidate terms using sparsity-promoting identification.

Neglecting structural lag states, an alternative and more general formulation is a second-order nonlinear dynamical system in which the aerodynamic-force acceleration at time step $n$ can contain any nonlinear combination of the aerodynamic $(Q_n,\dot{Q}_n)$ and structural $(u_n,\dot{u}_n, \ddot{u}_n)$ states:

\begin{equation}
\ddot{Q}_n = \mathcal{F}\!\left(Q_n,\dot{Q}_n, u_n,\dot{u}_n, \ddot{u}_n \right)
\label{eq:3}
\end{equation}

\noindent where $\mathcal{F}(\cdot)$ is to be discovered from a library of candidate terms using sparsity-promoting identification. This is proposed as a potential improvement to the lag-0 variant of Eq.~\ref{eq:4} (which does not contain cross-coupling between the aerodynamic and structural states). 

A second alternative to Eq.~\ref{eq:4} may be to combine $\mathcal{F}(\cdot)$ and $\mathcal{F}_S(\cdot)$, written as:

\begin{equation}
\ddot{Q}_n = \mathcal{F}\!\left(Q_n,\dot{Q}_n, u_n,\dot{u}_n, \ddot{u}_n \right)
+
\mathcal{F}_S\!\left(\bm{z}_n\right)
\label{eq:5}
\end{equation}

\noindent where $\mathcal{F}(\cdot)$ captures the dominant current (lag-0) fluid and fluid--structure dynamics, including nonlinear combinations and cross-couplings of $Q_n,\dot{Q}_n,u_n,\dot{u}_n,$ and $\ddot{u}_n$, and $\mathcal{F}_S(\cdot)$ is introduced to account for nonlinear unsteady memory effects that are known to be important in transonic shock-dominated flows.

Identifying $\mathcal{F}(\cdot)$,  $\mathcal{F}_F(\cdot)$ and $\mathcal{F}_S(\cdot)$ from data is a significant challenge and the remainder of this section describes the identification approaches used via a series of example problems.

        \subsection{Constructing the Nonlinear Oscillator with Memory Using Known Functions}
        
        It is assumed in this work that $\mathcal{F}_F(\cdot)$ could contain any nonlinear combinations of the state variables $\dot{Q}_n$ and $Q_n$ to phenomenologically describe the nonlinear flow oscillations. One such nonlinear dynamical system that was used by Dowell~\cite{dowell81}, among others, in bluff body aerodynamics is the Van der Pol oscillator. An alternative that will be used here, which is mathematically equivalent to the Van der Pol oscillator, is the Rayleigh oscillator: 

        \begin{equation}
        \label{eq:6}
        \ddot{Q}_n = \epsilon\left(1-\left(\frac{\dot{Q}_n}{\omega_FQ_{ref}}\right)^2\right)\dot{Q}_n  - {\omega_F}^2(Q_n + \bar{Q}) 
        \end{equation}

        \noindent where $Q_{ref} = {Q}_{max}-\bar{Q}$ defines the maximum deviation of the aerodynamic forces about the mean, $\bar{Q}$, and $\omega_F$ is the fluid LCO frequency which is the frequency of the shock oscillations in transonic buffet (or the wake frequency in the case of bluff body aerodynamics). The offset term $\bar{Q}{\omega_F}^2$ accounts for the non-zero static aerodynamic load. Eq.~\ref{eq:6} contains a negative linear damping term and a positive cubic damping term which are the two features that permit a self-excited LCO. 
        
        A simple extension that can capture the fluid-only oscillations and fluid–structure interactions is to add Parkinson’s galloping model~\cite{parkinson89} (including added-mass effects) to Eq.~\ref{eq:6}. The resulting equation becomes:
        
        \begin{equation}
        \label{eq:7}
        \ddot{Q}_n = \epsilon\left(1-\left(\frac{\dot{Q}_n}{\omega_FQ_{ref}}\right)^2\right)\dot{Q}_n  - {\omega_F}^2(Q_n + \bar{Q})  -B_1\ddot{u}_n + A_1\dot{u}_n - A_3\dot{u}_n^3 + A_5\dot{u}_n^5 - A_7\dot{u}_n^7
        \end{equation}

        \noindent {\color{black} where the Rayleigh-oscillator parameters $(\epsilon,\omega_F,\bar{Q},Q_{ref})$ and the coupling coefficients $(B_1,A_1,A_3,A_5,A_7)$ are identified from either wind-tunnel measurements or CFD time histories. The training data consists of a prescribed structural perturbation (rigid-body or elastic) and the corresponding generalized aerodynamic response. For buffet and other self-excited flows, it is also advantageous to include a \enquote{buffet-only} dataset with zero structural input, which helps in identifying the fluid-oscillator terms.} Note that $\dot{u}_n$ can be swapped for $u_n$ in the case of pitch motion. For a comprehensive and intuitive description of how this model is constructed, the reader is referred to Dowell~\cite{dowell81}. 

        To include nonlinear memory, the function $\mathcal{F}_S(\cdot)$ can be approximated using a $p^{th}$-order pruned Volterra series which, in general terms, for a causal, time-invariant, fading memory, nonlinear system, approximates an output $y_n$ due to an input $x_n$ as:

        \begin{equation}
        \label{eq:8}
        y_n = \sum_{j=1}^p \sum_{k=n^*}^{n} \mathcal{H}_{j,n-k} (x_k)^j 
        \end{equation}

        \noindent where $n^* = n-N_L$ and $\mathcal{H}_j$ is the main diagonal of the $j^{th}$-order pruned (diagonal) Volterra kernel which is unknown and must be identified. The galloping model in Eq.~\ref{eq:7} can be replaced with the pruned Volterra series approximation in Eq.~\ref{eq:8}. The resulting integro-differential equation (IDE) becomes:

        \begin{equation}
        \label{eq:9}
        \ddot{Q}_n = \epsilon\left(1-\left(\frac{\dot{Q}_n}{\omega_FQ_{ref}}\right)^2\right)\dot{Q}_n  - {\omega_F}^2(Q_n + \bar{Q})  -B_1\ddot{u}_n + \sum_{j=1}^p \sum_{k=n^*}^{n} \mathcal{H}_{j,n-k} (U_k)^j 
        \end{equation}

        \noindent where $U_k$ denotes the selected structural DOF sample at time index $k$. For heave-driven cases, $U_k=\dot{u}_k$ should be used, whereas for pitch-driven cases it is more appropriate to take $U_k=u_k$. If desired, both inputs can be included with separate kernel sets for $u_k$ and $\dot{u}_k$. Another important consideration is the equivalence between the galloping model coefficients in Eq.~\ref{eq:7} and the pruned Volterra kernels in Eq.~\ref{eq:9}, specifically: $A_1 \equiv  \mathcal{H}_{1,0}$, $A_3 \equiv \mathcal{H}_{3,0}$, and so on. The point is that the galloping model can be seen as a specific case of Eq.~\ref{eq:8} by setting (i) $N_L = 0$, (ii) $p = 7$, and (iii) neglecting the even-ordered terms.  

        The next subsection will show how the coefficients of these known functions can be identified from data, then go on to demonstrate how sparsity-promoting algorithms can be used to identify new equations $\mathcal{F}(\cdot)$ and $\mathcal{F}_F(\cdot)$ when given a large library of potential terms. 

        \subsection{Function Identification}
        \label{sec:funcID}

        {\color{black}This section describes a single-input single-output (SISO) identification strategy, appropriate because the present study considers only s-DOF aeroelastic systems (isolated heave or isolated pitch). Extension to multi-DOF aeroelasticity may require a multi-input formulation and training data in which multiple structural coordinates are excited simultaneously, enabling identification of nonlinear cross-coupling terms~\cite{candon25a}.}

        \subsubsection{Generation of Training Data}
        {\color{black}Training data are generated using unsteady CFD simulations at a known buffeting condition.} Each structural mode is excited in isolation using band-limited noise, $\bm{u}^i \in \mathbb{R}^N$, where $N$ is the number of samples. $\bm{u}^i$ is set to zero for the first $N_{B}$ time steps so that the training data contains a few cycles of the unsteady aerodynamic force oscillations due to buffet only. The generalized forces are projected onto each $j^{th}$ structural mode, given by $\bm{Q}^j \in \mathbb{R}^{N}$. For the remainder of this discussion, the aerodynamic force vector is referred to as $\bm{Q}$, noting that it could be related to any generalized force of interest, and generalized displacement as $\bm{u}$. A low-pass filter is applied to $\bm{Q}$ to remove high-frequency noise components from the data (specifics are given in the results Section~\ref{sec:res}). The aerodynamic and structural derivatives are then obtained using finite differences, and the full set of state variables are given as:
        
        \begin{equation}
        \label{eq:10}
            \begin{split}
                & \bm{u} := [u_0, \dots, u_{N-1}]\in \mathbb{R}^{N}, \quad \dot{\bm{u}} := [\dot{u}_0, \dots, \dot{u}_{N-1}]\in \mathbb{R}^{N}, \quad \ddot{\bm{u}} := [\ddot{u}_0, \dots, \ddot{u}_{N-1}]\in \mathbb{R}^{N},\\
                & \bm{Q} := [Q_0, \dots, Q_{N-1}]\in \mathbb{R}^{N}, \quad \dot{\bm{Q}} := [\dot{Q}_0, \dots, \dot{Q}_{N-1}]\in \mathbb{R}^{N}
            \end{split}
        \end{equation}

        The output variable is stored as $\ddot{\bm{Q}} := [\ddot{Q}_0, \dots, \ddot{Q}_{N-1}]\in \mathbb{R}^{N}$. 

        \subsubsection{Orthogonal Matching Pursuit}

        Orthogonal Matching Pursuit (OMP) is a greedy algorithm used to obtain a sparse approximation of a signal. Given a data (or dictionary) matrix $\bm{A} \in \mathbb{R}^{s \times d}$, a target vector $\bm{b} \in \mathbb{R}^s$, and a desired sparsity level $\kappa$, OMP seeks a solution $\bm{x} \in \mathbb{R}^d$ to the linear system: 
        
        \begin{equation}
            \bm{b} \approx \bm{A x}
        \end{equation}
        
        \noindent such that $\bm{x}$ has at most $\kappa$ nonzero entries. Formally, OMP aims to solve:
        \begin{equation*}
            \min_{x}\| \bm{b} - \bm{Ax} \|_2^2 
            \quad \text{subject to} \quad \|\bm{x}\|_0 \leq \kappa,
        \end{equation*}
        where $\|\bm{x}\|_0$ denotes the number of nonzero entries in $\bm{x}$. The method proceeds iteratively: at each iteration, it selects the column of $A$ most correlated with the current residual, adds that column to the active set, and recomputes the least-squares solution on the active set. This process continues until $\kappa$ terms have been selected or the residual becomes sufficiently small. 

        \subsubsection{Coefficient Estimation of Known Functions}
        \label{sec:knownFunc}
        
        Initially two relatively straightforward examples are considered: estimating the coefficients of the Rayleigh oscillator (Eq.~\ref{eq:6}) and the Rayleigh-Parkinson equation (Eq.~\ref{eq:7}). This requires the construction of the state matrices:

        \begin{equation}
        \label{eq:11}
        \bm{\Phi}_r = [\bm{\dot{Q}}, \bm{\dot{Q}}^3, \bm{Q}, \mathbb{J} ] \in \mathbb{R}^{N \times 4}
        \end{equation}

        \begin{equation}
        \label{eq:12}
        \bm{\Phi}_{rp} = [\bm{\dot{Q}}, \bm{\dot{Q}}^3, \bm{Q}, \mathbb{J}, \bm{\ddot{u}}, \bm{\dot{u}}, \bm{\dot{u}}^3,\bm{\dot{u}}^5, \bm{\dot{u}}^7] \in \mathbb{R}^{N \times 9}
        \end{equation}
        
        \noindent where $\mathbb{J} = [ \text{---} 1 \text{---}]^T \in \mathbb{R}^{N}$. It should be recalled here that $\dot{u}$ can be replaced by $u$ if needed. The coefficients are computed by solving the least-squares problems:
        
        \begin{equation*}
        \label{eq:13}
        \min_{\bm{c}_{r}}\|\bm{\Phi}_{r} \bm{c}_{r}-\bm{\ddot{Q}}\|_2^2, \quad  \quad \min_{\bm{c}_{rp}}\|\bm{\Phi}_{rp} \bm{c}_{rp}-\bm{\ddot{Q}}\|_2^2
        \end{equation*}

        \noindent which yields the solutions:
        
        \begin{equation}
        \label{eq:14}
        \bm{c}_{r} = \bm{\Phi}_{r}^+\bm{\ddot{Q}} = [c_{r1}, \hdots, c_{r4}]^T \in \mathbb{R}^{4}, \quad \quad  \bm{c}_{rp} = \bm{\Phi}_{rp}^+\bm{\ddot{Q}} = [c_{rp1}, \hdots, c_{rp9}]^T \in \mathbb{R}^{9}
        \end{equation}
        
        \noindent where $^+$ is the Moore-Penrose pseudoinverse. This identifies the coefficients of Eq.~\ref{eq:6} ($\bm{c}_r$) and Eq.~\ref{eq:7} ($\bm{c}_{rp}$), giving the Rayleigh ROM ($\text{\textbf{ODE-ROM}}_R$) and Rayleigh-Parkinson ROM ($\text{\textbf{ODE-ROM}}_{RP}$):

        \begin{equation}
        \label{eq:15}
            \text{\textbf{ODE-ROM}}_R:  \ddot{Q}_n = c_{r1}\dot{Q}_n + c_{r2}\dot{Q}_n^3 +  c_{r3}Q_n  + c_{r4} 
        \end{equation}

        \begin{equation}
        \label{eq:16}
            \begin{split}
            \text{\textbf{ODE-ROM}}_{RP}:  \ddot{Q}_n =  &  \, c_{rp1}\dot{Q}_n + c_{rp2} \dot{Q}_n^3 +  c_{rp3}Q_n  + c_{rp4}  +  \\ &  c_{rp5}\ddot{u}_n + c_{rp6}\dot{u}_n + c_{rp7}\dot{u}_n^3 + c_{rp8}\dot{u}_n^5 + c_{rp9}\dot{u}_n^7 
            \end{split}
        \end{equation}

        \noindent where Eq.~\ref{eq:15} and Eq.~\ref{eq:16} contain all terms necessary to derive Eq.~\ref{eq:6} and Eq.~\ref{eq:7} respectively. 
        
        To identify the coefficients of Eq.~\ref{eq:9} (replacing the galloping model with a pruned Volterra series) a lower-left triangular circulant matrix of $\bm{U}:= [U_0, \hdots U_{N-1}]$ is constructed (recalling that $\bm{U} = \bm{u}$ or $\bm{U} = \dot{\bm{u}}$):

        \begin{align} 
        \label{eq:19}
        \bm{L} = \begin{bmatrix}
        U_0 & 0 & \hdots & 0\\
        U_1 & U_0 & \hdots & 0\\
        \vdots  &\vdots  &\vdots &  0\\
        U_{N-1}   & U_{N-2}   &\hdots & U_{N-1-N_L}\\
          \end{bmatrix}\in \mathbb{R}^{N \times (N_L+1)}
          \end{align}

          \noindent which, for a pruned Volterra series up to order $p$, replaces the nonlinear structural terms in the state matrix as follows:

        \begin{equation}
        \label{eq:20}
        \bm{\Phi}_{rv} = [\bm{\dot{Q}}, \bm{\dot{Q}}^3, \bm{Q}, \mathbb{J}, \bm{\ddot{u}}, \bm{L}, \bm{L}^2, \hdots, \bm{L}^p] \in \mathbb{R}^{N \times N_I}
        \end{equation}

        \noindent where the number of IDE terms $N_I = 5+p(N_L+1)$. While it is possible to solve this using a standard least-squares optimization, identifying all the pruned Volterra series terms is likely unnecessary and Orthogonal Matching Pursuit (OMP) is used to solve the $\ell_0$-minimization problem:

        \begin{equation*}
        \label{eq:21}
        \min_{\bm{c}_{rv}}\|\bm{\Phi}_{rv} \bm{c}_{rv}-\bm{\ddot{Q}}\|_2^2 \quad \text{subject to} \quad \| \bm{c}_{rv} \|_0 < \kappa
        \end{equation*}

        \noindent where $\kappa$ is the pre-defined number of non-zero coefficients. A grid search of $\kappa$ and the number of lags, $N_L$, is conducted, and the optimized sparse set of coefficients is obtained:
        
        \begin{equation}
        \label{eq:22}
        \bm{c}_{rv} = \text{OMP}(\bm{\Phi}_{rv}, \bm{\ddot{Q}}, \kappa, \text{constraints}) = [c_{rv1}, \hdots, c_{rv5}, \mathcal{H}_{1}, \hdots, \mathcal{H}_{p}]^T \in \mathbb{R}^{N_I}
        \end{equation}

        \noindent and it follows that the Rayleigh-Volterra IDE ROM ($\text{\textbf{IDE-ROM}}_{RV}$) is given by:

        \begin{equation}
        \label{eq:23}
            \text{\textbf{IDE-ROM}}_{RV}:  \ddot{Q}_n =  c_{rv1}\dot{Q}_n + c_{rv2} \dot{Q}_n^3 +  c_{rv3}Q_n  + c_{rv4} + c_{rv5}\ddot{u}_n + \sum_{j=1}^p \sum_{k=n^*}^n \mathcal{H}_{j,n-k}(U_k)^j 
        \end{equation}
        
        \subsubsection{Identification of Unknown Functions}

        In this section, the process for discovering new equations is described. The first step is to create a linear state matrix as follows:

        \begin{equation}
        \label{eq:26}
        \bm{\Phi}_{D_l} = [\bm{\dot{Q}}, \bm{Q}, \bm{\ddot{u}}, \bm{\dot{u}}, \bm{u}] \in \mathbb{R}^{N \times 5}
        \end{equation}

        \noindent and polynomial features up to order $p_{o}$ are added, giving the nonlinear state matrix: 

        \begin{equation}
            \label{eq:27}
            \bm{\Phi}_{D_{nl}} = [\bm{\dot{Q}}, \bm{Q}, \bm{\ddot{u}}, \bm{\dot{u}}, \bm{u}, \bm{\dot{Q}}^2, \bm{\dot{Q}}\bm{Q}, \bm{\dot{Q}}\bm{\ddot{u}}, \hdots, \bm{u}^{p_{o}}, \mathbb{J} ] \in \mathbb{R}^{N \times N_O}
        \end{equation}

        \noindent where $N_O = \binom{5+p_{o}}{p_{o}}$ is the total number of candidate terms and $\bm{\Phi}_{D_{nl}}$ contains an abundance of terms to formulate the unknown nonlinear equation. Setting the cardinality as the stopping criterion, the $\ell_0$-minimization problem is:

        \begin{equation*}
        \label{eq:28}
        \min_{\bm{c}_{D}}\|\bm{\Phi}_{D_{nl}} \bm{c}_{D}-\bm{\ddot{Q}}\|_2^2 \quad \text{subject to} \quad \| \bm{c}_{D} \|_0 < \kappa
        \end{equation*}
        
        \noindent which is solved using OMP to obtain the coefficients of the discovered ODE ROM ($\text{\textbf{ODE-ROM}}_{D}$) as follows:

        \begin{equation}
        \label{eq:29}
        \text{\textbf{ODE-ROM}}_{D}: \bm{c_{D}} = \text{OMP}(\bm{\Phi}_{D_{nl}},\bm{\ddot{Q}}, \kappa, \text{constraints}) = [c_{D1}, \hdots, c_{DN_O}]^T \in \mathbb{R}^{N_O}
        \end{equation}

        \noindent of which the cardinality $\kappa = \|\bm{c}_{D}\| << N_O$, and $\text{supp}(\bm{c}_{D})$ can be used to extract the symbolic equation. 
        
        The buffet-only oscillator model can be discovered by removing the structural states from Eq.~\ref{eq:26}, yielding the nonlinear state matrix: 
        
        \begin{equation}
            \label{eq:30}
            \bm{\Phi}_{DB_{nl}} = [\bm{\dot{Q}}, \bm{Q}, \bm{\dot{Q}}^2, \bm{\dot{Q}}\bm{Q}, \hdots, \bm{Q}^{p_{o}}, \mathbb{J} ] \in \mathbb{R}^{N \times \binom{2+p_o}{p_o}}
        \end{equation}
        
        \noindent and the OMP solution to the $\ell_0$-minimization problem yields the coefficients of the discovered buffet-only ODE ROM  ($\text{\textbf{ODE-ROM}}_{DB}$):

        \begin{equation}
        \label{eq:31}
        \text{\textbf{ODE-ROM}}_{DB}: \bm{c}_{DB} = \text{OMP}(\bm{\Phi}_{DB_{nl}},\bm{\ddot{Q}},\kappa, \text{constraints}) = [c_{DB_1}, \hdots, c_{DB_{N_O}}] \in \mathbb{R}^{N_O}
        \end{equation}
        
        The pruned Volterra series terms are added by expanding the state matrix $\bm{\Phi}_{D_{nl}}$ (Eq.~\ref{eq:27}) in the same way as previously described:

        \begin{equation}
            \label{eq:32}
            \bm{\Phi}_{DV_{nl}} = [\bm{\dot{Q}}, \bm{Q}, \bm{\ddot{u}}, \bm{\dot{u}}, \bm{u}, \bm{\dot{Q}}^2, \bm{\dot{Q}}\bm{Q}, \bm{\dot{Q}}\bm{\ddot{u}}, \hdots, \bm{u}^{p_{o}}, \mathbb{J} , \bm{L}, \bm{L}^2, \hdots, \bm{L}^p] \in \mathbb{R}^{N \times N_I}
        \end{equation}

        \noindent where the number of candidate terms is $N_I = \binom{5+p_{o}}{p_{o}} + p(N_L+1)$. In identification, the terms identified in $\bm{c}_{D}$ are fixed (and set all others to zero) while allowing new coefficients for those terms to be identified (along with the pruned Volterra series terms). The constrained $\ell_0$-minimization problem is therefore defined as:

        \begin{equation}
        \label{eq:33}
        \min_{\bm{c}_{DV}}
        \left\| \bm{\Phi}_{DV_{nl}} \bm{c}_{DV} - \ddot{\bm{Q}} \right\|_2^2
        \quad \text{subject to} \quad
        \|\bm{c}_{DV}\|_0 < \kappa,
        \quad
        \mathrm{supp}\!\left(\bm{c}_{DV}^{(ODE)}\right) \subseteq \mathrm{supp}\!\left(\bm{c}_{D}\right),
        \end{equation}

        \noindent where $\bm{c}_{DV}=\big[(\bm{c}_{DV}^{(ODE)})^T\ (\bm{c}_{DV}^{(V)})^T\big]^T$, with $\bm{c}_{DV}^{(ODE)}\in\mathbb{R}^{N_O}$ and $\bm{c}_{DV}^{(V)}\in\mathbb{R}^{p(N_L+1)}$ the Volterra coefficients. This is solved using OMP to obtain the coefficients of the discovered IDE ROM ($\text{\textbf{IDE-ROM}}_{D}$):

        \begin{equation}
        \label{eq:34}
        \text{\textbf{IDE-ROM}}_{D}: \bm{c}_{DV} = \text{OMP}(\bm{\Phi}_{DV_{nl}}, \bm{\ddot{Q}}, \kappa, \text{constraints}) = [c_{DV_1}, \hdots, c_{DV_{N_O}}, \mathcal{H}_{DV_1}, \hdots, \mathcal{H}_{DV_p}] \in \mathbb{R}^{N_I}
        \end{equation}
        
        \noindent where again $\kappa << N_I$. Table~\ref{tab:roms} summarizes how each of the ROMs that have been described in this section are constructed. 

        \begin{table}[h]
        \caption{Summary of reduced-order model differential equations.}
        \label{tab:roms}
        \begin{tabular}{ccccc}
        \hline
            \hline
             & fluid oscillator & structural terms & ID alg. & constraints \\
            \hline
            $\text{\textbf{ODE-ROM}}_{R}$ & Rayleigh & -  & LS & -   \\
            $\text{\textbf{ODE-ROM}}_{DB}$ & Disc. ODE & -  & OMP & -   \\
            $\text{\textbf{ODE-ROM}}_{RP}$ & Rayleigh & Parkinson  & LS & -   \\
            $\text{\textbf{IDE-ROM}}_{RV}$ & Rayleigh &  Volterra & OMP & -  \\
            $\text{\textbf{ODE-ROM}}_{D}$ & Disc. ODE & Disc. ODE  & OMP & -  \\
            $\text{\textbf{IDE-ROM}}_{D}$ & Disc. ODE & Disc. ODE + Volterra & OMP & fluid + struct. ODE terms \\
            \hline
            \hline
        \end{tabular}
    
        \end{table}

        {\color{black}As a final note, variants of the ROMs that enforce hard constraints on the fluid-only terms and coefficients were also investigated. In the present setting, hard constraints were found to over-constrain the model and degrade the accuracy of the fluid--structure coupling. Soft constraints ($e.g.$, penalty terms) may alleviate this issue and represent a promising extension.}

    \section{Computational Framework}
    
     {\color{black}The present study has been performed for the ONERA OAT15A airfoil, with experimental measurements (unsteady aerodynamic only) available from the transonic wind tunnel of the Onera-Meudon Centre in France~\cite{jacquin09}. The experimental model is designed to study two-dimensional buffet on a rigid wing, with a chord length of $c = 0.23$m, a span of 0.78m and a blunt trailing edge of thickness 0.005$c$.} Experiments have been performed over a Mach number range of $0.70 \leq M_\infty \leq 0.75$ over a wind-off angle-of-attack (AOA) sweep of $2.4^\circ \leq \alpha_0 \leq 3.91^\circ$ to determine the transonic buffet envelope onset at a Reynolds number of $Re_\infty = 3\times10^6$ (based on the chord length).

    \subsection{Computational Fluid Dynamics Model}
    The general-purpose finite volume code ANSYS Fluent 2024 R2~\cite{ansys} is used and two separate computational models are considered. Computational Model 1 (CM1) considers a finer spatially converged grid for the purpose of verifying the system identification approach across a range of static AOAs for the buffet-only case. Computational Model 2 (CM2) is designed for efficiency in the aeroelastic component of the paper. The model considers less spatial and temporal resolution. Given that this paper is concerned with reduced-order modeling (where the CFD model is considered to be ground truth) detailed temporal and spatial convergence studies are not considered.
    
    \subsubsection{Computational Model 1: Aerodynamics Only}
     For Computational Model 1 (CM1), the URANS equations are solved using a coupled pressure-based solver. Convective terms are discretized with an implicit second-order upwind scheme, with Rhie–Chow distance-based flux interpolation, while diffusive terms use second-order central differencing. A dual time-stepping scheme is used with bounded second-order implicit temporal discretization. A non-dimensional time-step of $\Delta \tau = \Delta t(u_\infty/c) = 5\times 10^{-3}$ is used which gives a temporal resolution of approximately 200 steps per convective time unit (CTU). The computational grid (Fig.~\ref{cm1}) is a structured C-grid topology with one cell in the spanwise direction, as provided for the DPW-8/AePW-4 Buffet Working Group~\footnote{https://aiaa-dpw.larc.nasa.gov/grids.html}. The average non-dimensional first cell height is $\bar{y}^+ = 0.179$. 
     
    \subsubsection{Computational Model 2: Aeroelastic}
\begin{figure}[h]
		\centering
            \subfigure[CM1 computational grid (118,300 cells)]{\label{cm1}
			\includegraphics[width=0.45\textwidth]{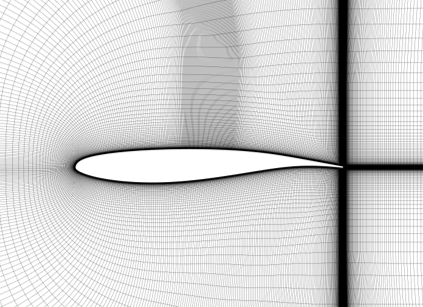}}
            \subfigure[CM2 computational grid (47,400 cells)]{\label{cm2}
			\includegraphics[width=0.45\textwidth]{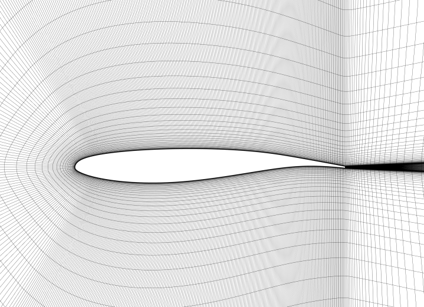}}
		\caption{Computational grids}
		\label{fig:mesh}
	\end{figure}

     For Computational Model 2 (CM2), the two-dimensional URANS equations are solved using the density-based implicit solver with second-order upwind Roe-flux splitting scheme for the advective terms, and central-differencing for the diffusive terms. A dual time-stepping scheme is employed with second-order implicit temporal discretization and with a non-dimensional time-step of $\Delta \tau = 1\times 10^{-2}$ (100 steps per CTU). The computational grid (Fig.~\ref{cm2}) is a structured C-grid topology with forced transition (represented by separate domains) imposed at ($x/c$) = 0.07. The average non-dimensional first cell height is $\bar{y}^+ = 0.94$.

\begin{figure}[h]
		\centering
			\includegraphics[width=1\textwidth]{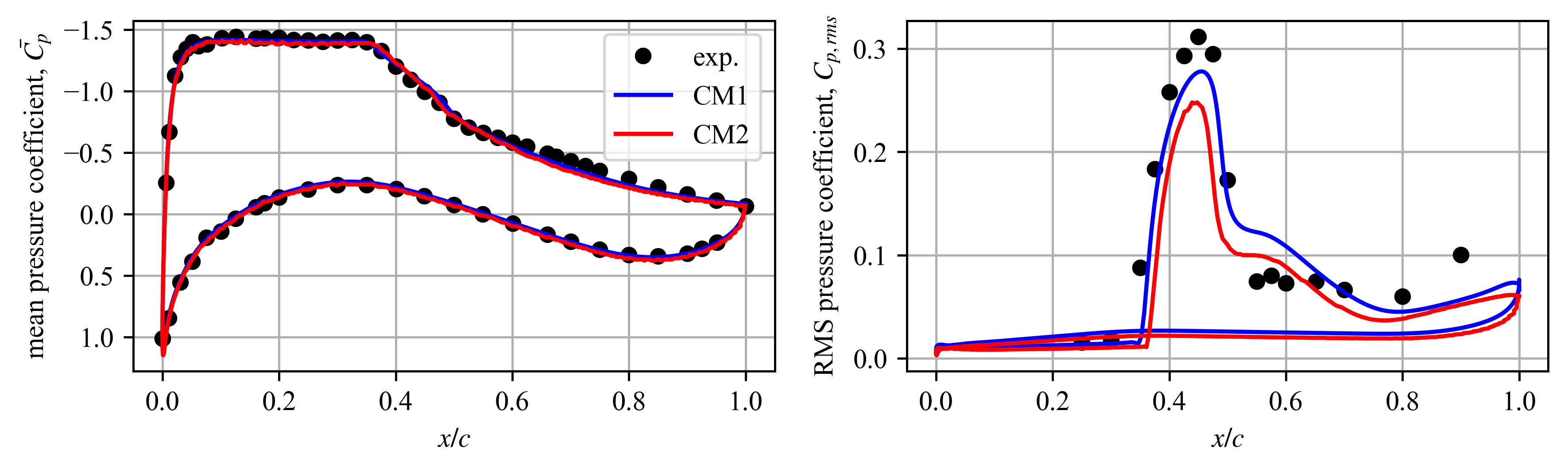}
		\caption{Mean and RMS pressure coefficient compared to experimental results for CM1 and CM2. }
		\label{fig:validation}
	\end{figure}

     Both models use the SST $k-\omega$ turbulence model with curvature correction~\cite{spalart97} and have convergence criteria set to $1\times10^{-5}$ for the scaled residuals at each time step. Validation of the two models is presented in Fig.~\ref{fig:validation}, demonstrating sufficient accuracy for the purposes of this research. Spatial and temporal refinement for CM1 is presented by Candon \textit{et al.}~\cite{candon25e} and for CM2 by Carrese \textit{et al.}~\cite{carrese16}.  

    \subsection{Aeroelastic Equation of Motion}
    In this work s-DOF structural equations of motion are considered separately in the heave, $h$, and pitch, $\alpha$, modes, given as:
        \begin{equation}
        \label{eq:35}
            m(\ddot{h} + 2\zeta_h \omega_h \dot{h} + {\omega_h}^2h) = q_\infty c C_L
        \end{equation}

    \begin{equation}
    \label{eq:36}
        I_\alpha(\ddot{\alpha} + 2\zeta_\alpha \omega_\alpha \dot{\alpha} + {\omega_\alpha}^2\alpha) = q_\infty c^2 C_M
    \end{equation}
    
     \noindent where $m$ is the sectional mass, $\omega_h$ and $\omega_\alpha$ are the heave and pitch natural frequencies respectively, and $\zeta_h$ and $\zeta_\alpha$ are the structural damping ratios. The sectional moment of inertia, $I_\alpha = \mu \pi \rho_\infty b^4 r_\alpha^2$, where the baseline structural-to-fluid mass ratio $\mu = m/(\pi\rho b^2) = 870$, and $r_\alpha^2 = 0.75$. The semi-chord, $b=0.115$m and the freestream fluid density, $\rho = 0.923$kg/m$^3$. These parameters are selected based on the work of Giannelis \textit{et al.}~\cite{giannelis16}. The elastic axis is located at $x/c = 0.25$. The generalized forces $C_L$ and $C_M$ are the lift and moment (about quarter-chord) coefficients respectively, and $q_\infty$ is the freestream dynamic pressure.  
     
     {\color{black}The structural equations of motion are integrated forward in time both within the CFD solver (to provide a high-fidelity aeroelastic reference) and within the reduced-order modeling framework. In both cases, the generalized forces required by the structural equations are evaluated at each time step. In the CFD/CSD simulations, the structural ODEs are implemented in ANSYS Fluent using a user-defined function (UDF) and advanced in time using a fourth-order Runge–Kutta scheme. The associated flow–mesh motion is handled using Fluent’s dynamic-mesh formulation with diffusive smoothing, which maintains near-wall mesh quality while distributing deformation into the far field. In the ROM-based simulations, the coupled aeroelastic system is integrated using an explicit Euler scheme.}
     
        \clearpage
     \section{Results}
     \label{sec:res}

     In this section results are presented for the buffet-only and aeroelastic reduced-order models. Unless otherwise stated, baseline operating conditions are used which consider a freestream Mach number $M_\infty = 0.73$, Reynolds number $Re_\infty \approx 3\times10^6$ (based on the chord length), and wind-off AOA, $\alpha_0 = 3.5^\circ$. 
    
     \subsection{Buffet-Only Reduced-Order Model Differential Equations}
\begin{figure}[h]
        \centering
            \includegraphics[width=1\textwidth]{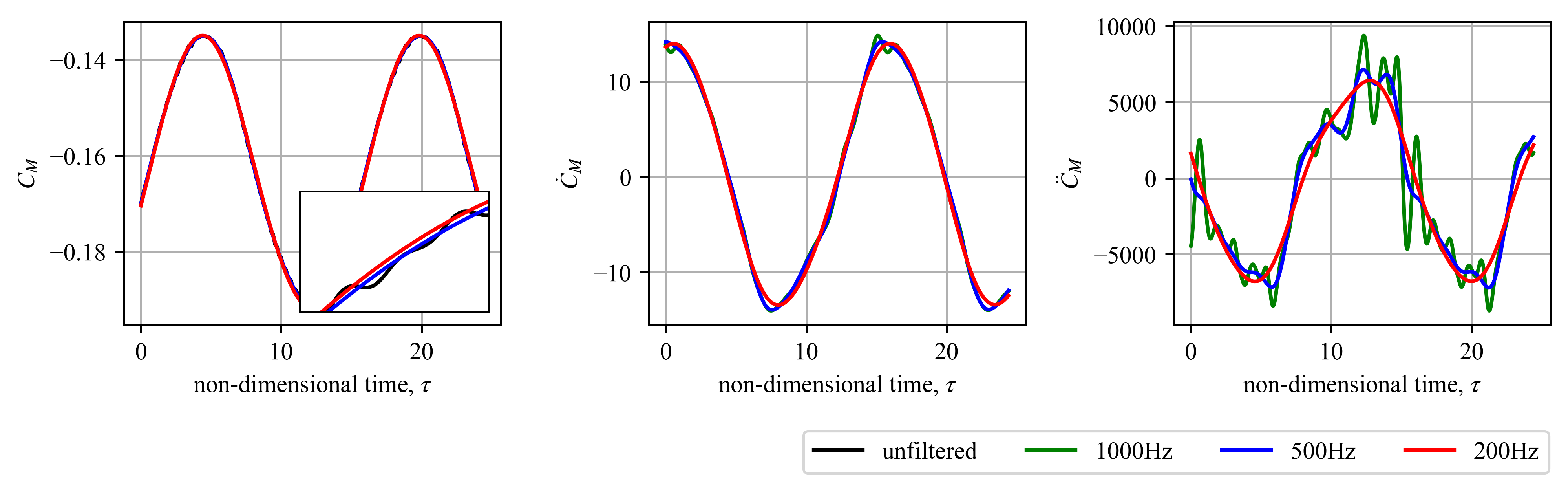}
        \caption{Influence of low-pass filtering on the moment coefficient at $\alpha_0 = 3.5^\circ$}
        \label{fig:filter}
    \end{figure}

     This section considers the identification of differential equation ROMs for the fluid-only LCO (in the absence of structural motion), primarily to verify the identification strategy. {\color{black}The aerodynamic forces are low-pass filtered with a 300Hz cutoff. This retains the buffet fundamental and first three harmonics ($\omega_B$, $2\omega_B$, $3\omega_B$, $4\omega_B$) and, for the highest structural natural frequency of interest, it retains the natural frequency and its first harmonic ($\omega_n$, $2\omega_n$). It is assumed that this is sufficient to model the buffet-driven aeroelastic responses where the mechanism of lock-in is primarily driven by interactions between the dominant buffet frequency and the structural natural frequency. The influence of filtering on the moment coefficient can be observed in Fig.~\ref{fig:filter} where the primary effect is to reduce noise in the derivatives.}

    \subsubsection{Rayleigh Oscillator Models}
    In this section the Rayleigh differential equation ROM, $\text{\textbf{ODE-ROM}}_R$, is identified for a sweep of wind-off AOAs using CM1. The $\text{\textbf{ODE-ROM}}_R$ formulation contains only the Rayleigh oscillator terms, forcing them to be identified. The training signal at each AOA contains approximately ten cycles of the buffet response once a stable LCO is achieved (growth/decay of the aerodynamic force is neglected). The objective is to assess whether the coefficients can be accurately and consistently identified, and that the time integrated Rayleigh oscillator model can reasonably reproduce the buffet response. Considering the four coefficients of the Rayleigh oscillator in Eq.~\ref{eq:6}, it can be re-written as: 

    \begin{equation}
        \label{eq:res1}
        \ddot{Q} - \epsilon\left(1-\left(\frac{\dot{Q}}{A}\right)^2\right)\dot{Q}  + B(Q + C) = 0
    \end{equation}

    \noindent where the generalized aerodynamic force $Q$ denotes the aerodynamic coefficient of interest (either $C_L$ or $C_M$), and the physical interpretation of the coefficients (denoted by $_{PH}$) is:

    \begin{equation*}
        \label{eq:res2}
        A_{PH} = \omega_B Q_{ref}, \quad  \quad B_{PH} = {\omega_B}^2, \quad  \quad C_{PH} = \bar{Q}, \quad \quad Q_{ref} = {Q}_{max}-\bar{Q}
    \end{equation*}

    \noindent and the identified interpretation from Eq.~\ref{eq:15} (denoted by $_{ID}$) is:

    \begin{equation*}
        \label{eq:res3}
        \epsilon = c_{r1}, \quad  \quad A_{ID} = \sqrt{-\frac{c_{r1}}{c_{r2}}}, \quad  \quad B_{ID} = -c_{r3}, \quad  \quad C_{ID} = \frac{c_{r4}}{c_{r3}}
    \end{equation*}

    With the sign convention in Eq.~\ref{eq:res1} and the coefficients identified in Eq.~\ref{eq:15}, it is required that $c_{r1}>0$ (giving negative linear damping), $c_{r2}<0$ (giving positive cubic damping), $c_{r3}<0$, and $c_{r4}$ may take either sign. If these sign conditions are not satisfied in a least-squares fit, the identified coefficients no longer correspond to a self-excited Rayleigh-type oscillator.
    
\begin{figure}[h]
		\centering
			\includegraphics[width=1\textwidth]{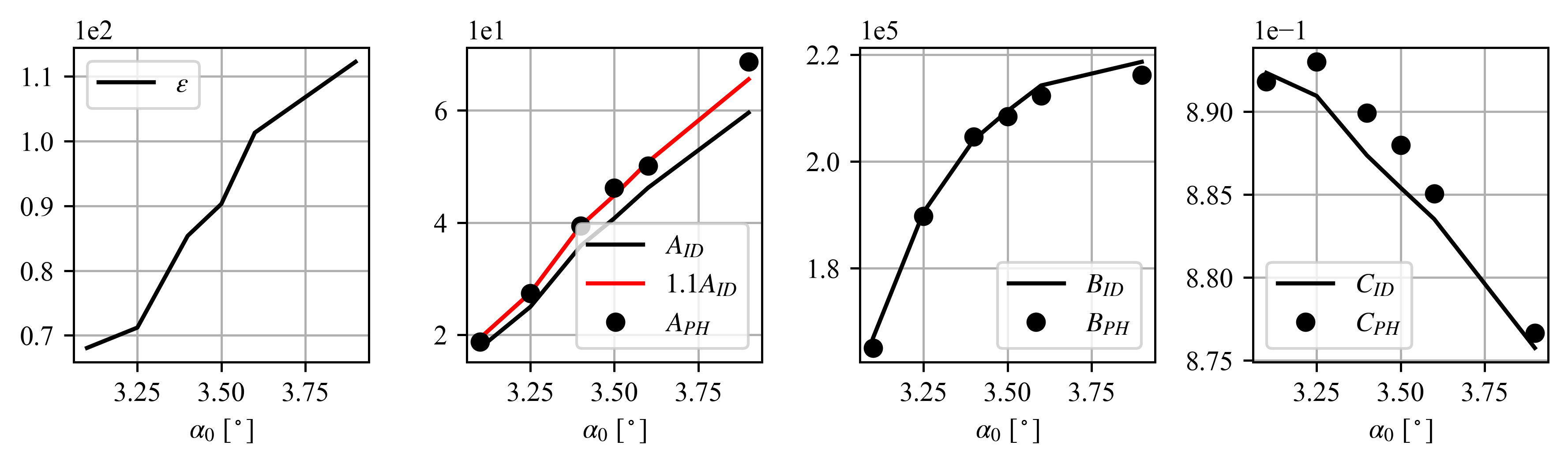}
		\caption{True and identified constants of the Rayleigh oscillator for lift force. }
		\label{fig:coefficients_lift}
	\end{figure}

\begin{figure}[h]
		\centering
			\includegraphics[width=1\textwidth]{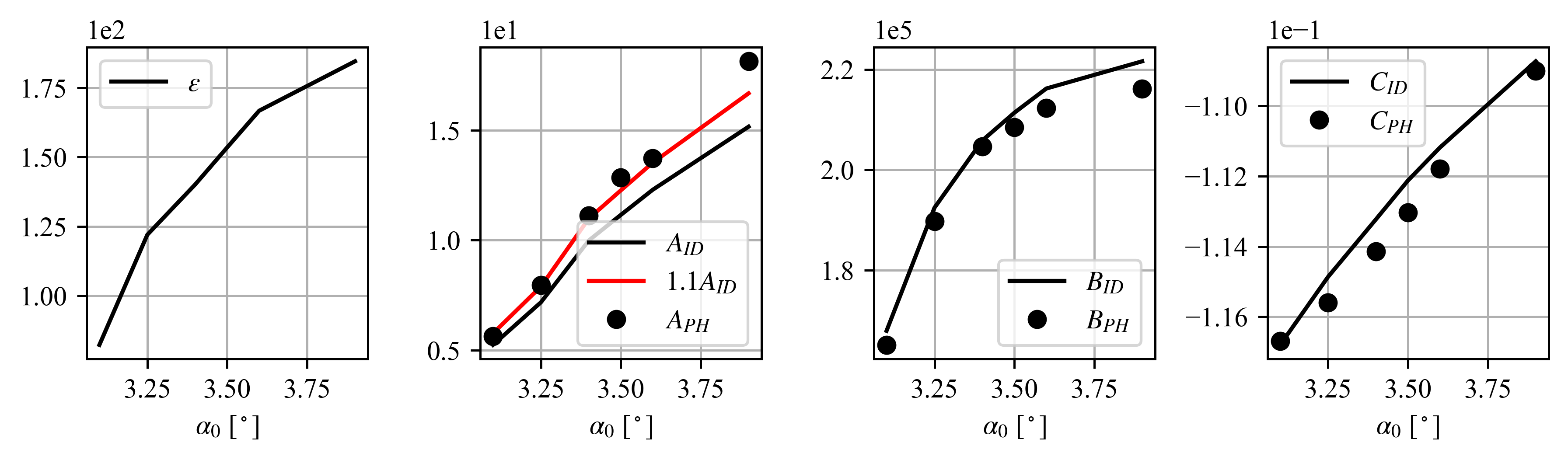}
		\caption{True and identified constants of the Rayleigh oscillator for pitching moment. }
		\label{fig:coefficients_moment}
	\end{figure}

\begin{figure}[h!]
		\centering
			\includegraphics[width=0.875\textwidth]{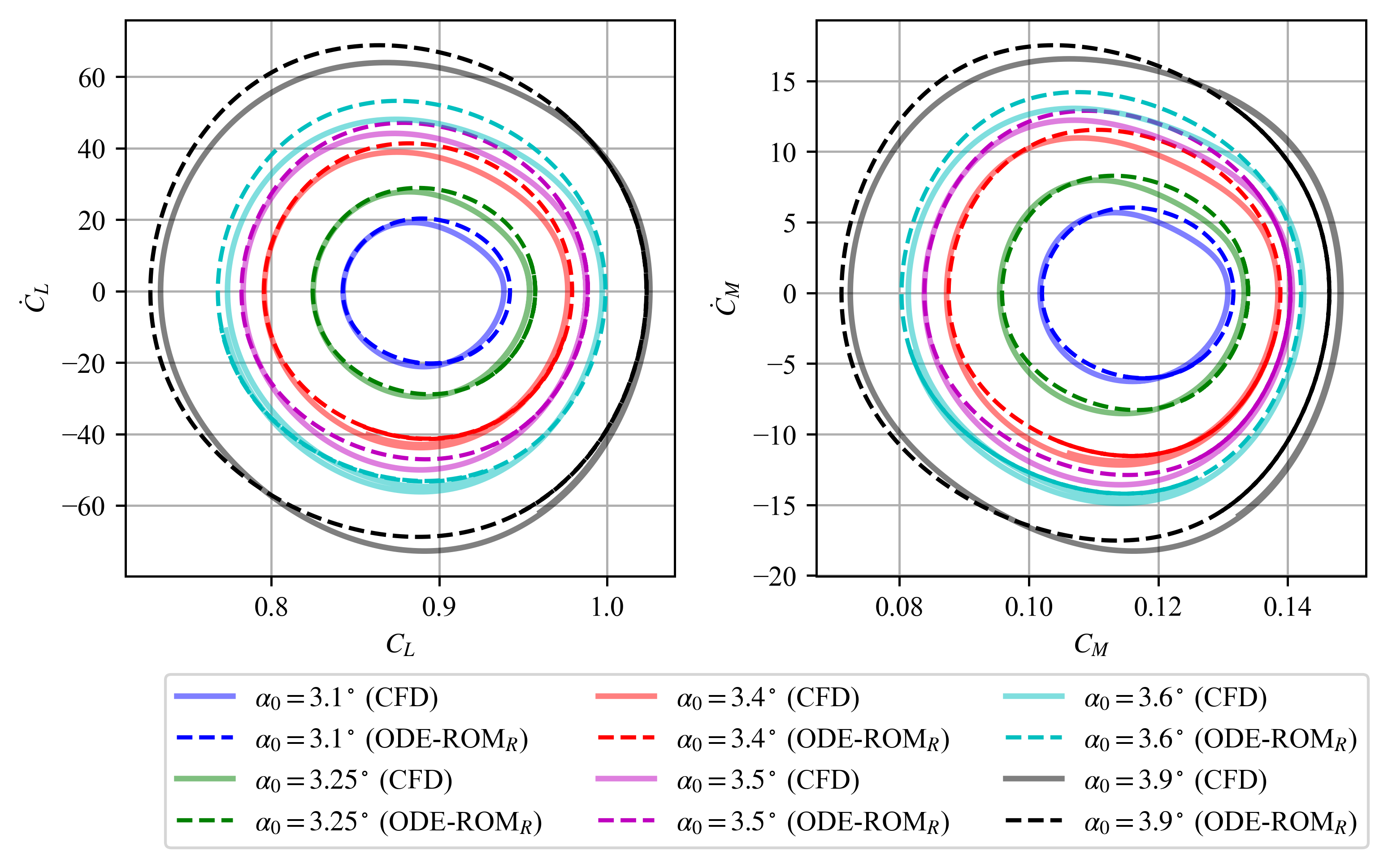}
		\caption{Comparison of buffet cycles computed via CFD (CM1) and the Rayleigh oscillator models.}
		\label{fig:rayleigh_crossval}
	\end{figure}

    Figures~\ref{fig:coefficients_lift} and~\ref{fig:coefficients_moment} present the identified and physical (true) values of these constants in the Rayleigh equation where it can be seen that for all AOAs the identified coefficients match the values of the physical coefficients well. The identified $A_{ID}$ terms, which define the maximum amplitude of the generalized force oscillations about the mean, are consistently approximately 10\% less than the physical value, $A_{PH}$, due to the mild nonlinear distortion. This can be accounted for by adding a scaling constant to $A$ in Eq.~\ref{eq:res1}. Phase portraits of the buffet response comparing the CFD result to the time-integrated Rayleigh equation are presented in Fig.~\ref{fig:rayleigh_crossval}. The identified Rayleigh oscillator models perform as expected, providing a good approximation of the mildly nonlinear buffet response. {\color{black}Although not shown, if an initial perturbation is applied, the trajectory rapidly decays or grows to the buffet limit cycle.}

    \subsubsection{Discovered Oscillator Models}
\begin{figure}[h]
        \centering
            \includegraphics[width=0.75\textwidth]{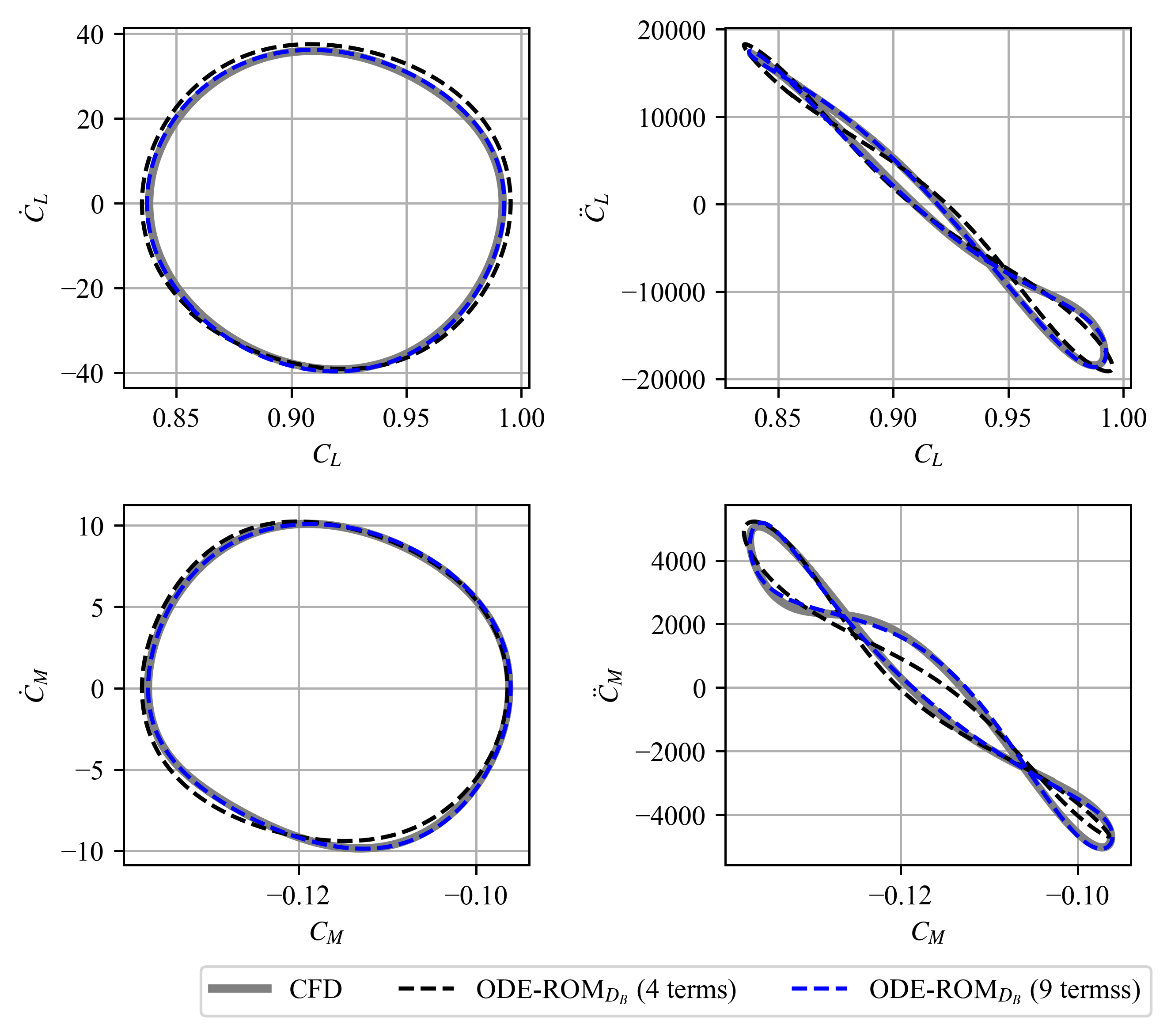}
        \caption{Comparison of buffet cycles computed via CFD (CM2) and the discovered oscillator models at $\alpha_0 = 3.5^\circ$.}
        \label{fig:discovered_buffet_crossval}
    \end{figure}

    Ordinary differential equation ROMs are now discovered for the buffet-only case, $\text{\textbf{ODE-ROM}}_{DB}$ using the CM2 model. Only the nominal wind-off $\alpha_0 = 3.5^\circ$ is considered. With prior knowledge that the Rayleigh oscillator is well suited, the nonlinear state matrix is constructed with only odd-ordered monomials up to order 5. Again, the training data considers approximately ten buffet cycles and the transient is neglected. Testing is completed by marching the discovered ODEs forward in time. Initially, the number of terms to be identified is set to $\kappa=4$ - the minimum number of terms required to obtain a nonlinear oscillator that will permit a self-excited stable LCO. The same ODE is identified for $C_L$ and $C_M$, as follows: 

    \begin{equation}
        \label{eq:res4}
        \ddot{Q} + c_{DB_1}\dot{Q}^5 + c_{DB_2}Q^4\dot{Q} + c_{DB_3}Q + c_{DB_4} = 0
    \end{equation}

    \noindent where the identified value $c_{DB_3} = {\omega_B}^2$, $c_{DB_1}>0$ and $c_{DB_2}<0$. This oscillator model, with negative fifth-order amplitude-dependent damping, and positive fifth-order damping behaves like the Rayleigh or Van der Pol oscillator, albeit with higher order terms. The number of terms is then increased incrementally until the oscillator model provides a near-exact fit. For both $C_L$ and $C_M$, the number of terms is stopped at $\kappa = 9$, returning the oscillator model:

    \begin{equation}
        \label{eq:res5}
        \ddot{Q} + c_{DB_1}\dot{Q} + c_{DB_2}Q\dot{Q}^2 + c_{DB_3}Q^4\dot{Q} + c_{DB_4}Q^3\dot{Q}^2 + c_{DB_5}Q^2\dot{Q}^3 + c_{DB_6}Q\dot{Q}^4 + c_{DB_7}Q + c_{DB_8} = 0
    \end{equation}

    \noindent where again the identified value $c_{DB_7} = {\omega_B}^2$. The time-integrated oscillator models compared to the CFD data are presented in Fig.~\ref{fig:discovered_buffet_crossval} where it can be seen that the nine-term oscillator model can nearly exactly model the buffet oscillations in both lift and moment.

    The results up to now are not entirely novel: other authors have shown that by using approaches based on sparse identification~\cite{sansica22,ma25}, oscillator models can be identified to describe buffet, and their coefficients can be interpolated (or cautiously extrapolated) to produce a reduced-order model. The next sections extend the approach to model the aeroelastic response which is the primary novel contribution of the paper.

    \subsection{Aeroelastic Training Input Signals}

\begin{figure}[h]
		\centering
            \subfigure[Time-domain]{\label{input_time}
			\includegraphics[width=1\textwidth]{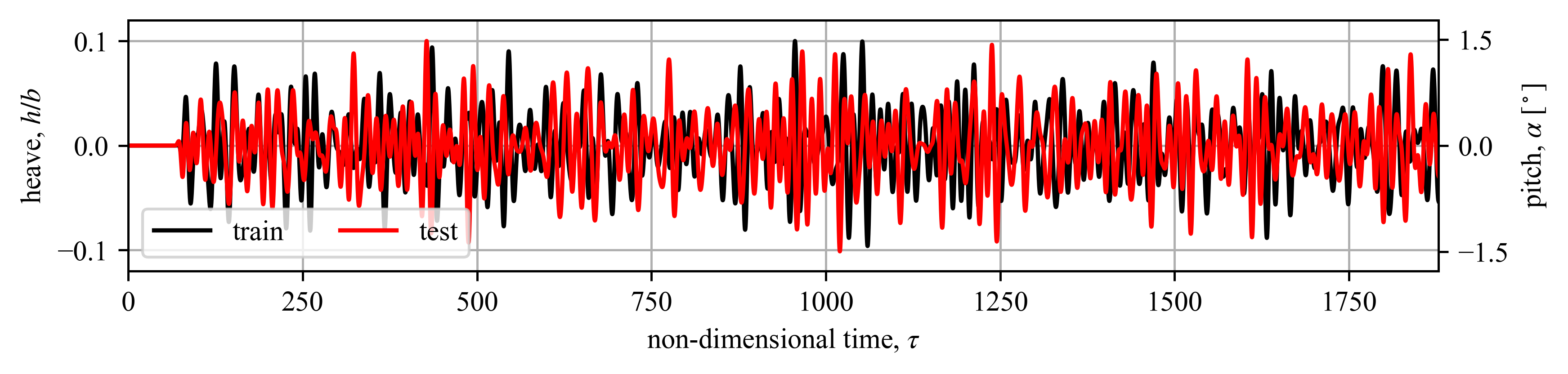}}
            
            \subfigure[Frequency-domain ]{\label{input_frequency}
			\includegraphics[width=1\textwidth]{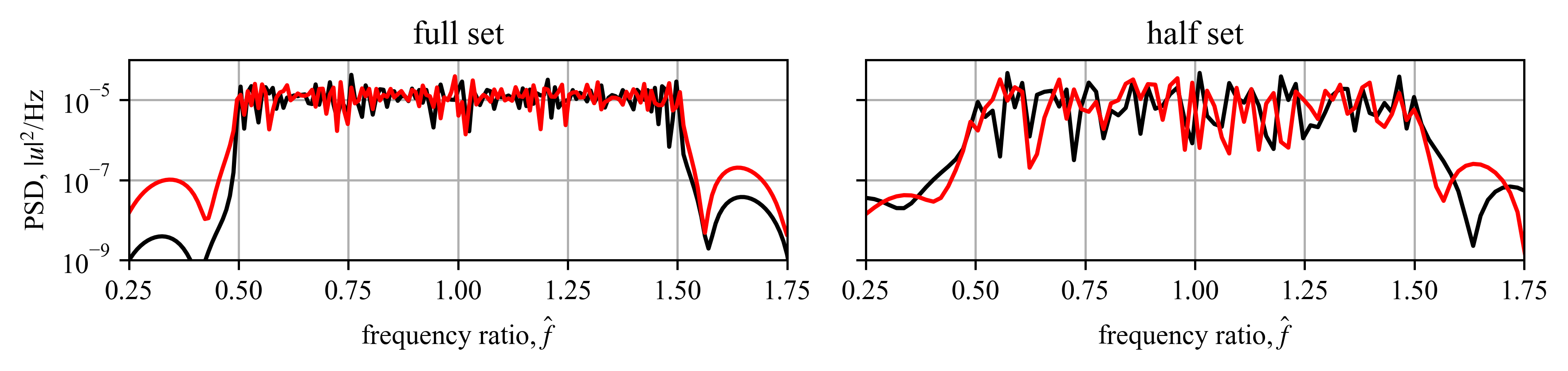}}
		\caption{Train and test input signals}
		\label{fig:train_inputs}
	\end{figure}

    Band-limited random excitation is used as an input signal to excite the system and record the aerodynamic response as presented in Fig.~\ref{fig:train_inputs}. The ROM is identified using the train signal and the test signal is reserved for cross-validation. {\color{black}The frequency band is $0.5 \leq \hat{f} \leq 1.5$ which is chosen based on the minimum and maximum structural natural frequencies of interest in the aeroelastic simulations.} Both the heave and pitch systems use the same base train / test signals (with different scaling) which is valid given that a multi-input ROM formulation is not considered. The heave signal is scaled such that the maximum amplitude of excitation is $(h/b)_{max}=0.1$, and pitch such that the maximum amplitude of excitation is $\alpha_{max}=1.5^\circ$. These frequency and amplitude ranges leave room for extrapolative capacity of the ROMs to be tested. The full set which is used for the studies of the pitch system contains 200k samples, the half set used for heave contains 100k samples. Although this may seem like a very large number of training samples, it is driven primarily by the small time step needed in shock buffet CFD models. Moreover, it is expected that this can be significantly reduced with well designed training inputs, however, it is outside of the scope of this study which aims only to propose the ROM methodology. Given the high cost of shock buffet aeroelastic simulation, the computational savings remain significant as will be shown later in the paper.  
    \clearpage

    \subsection{Aeroelastic Reduced-Order Model Differential Equations: Heave Motion}
    \label{sec:aeroelastic_heave}
    This section is concerned with modeling the s-DOF heave response of the airfoil under shock buffet oscillations.
    
    \subsubsection{Training and Cross-Validation}
    \label{sec:heave_train}
\begin{figure}[h!]
		\centering
			\includegraphics[width=1\textwidth]{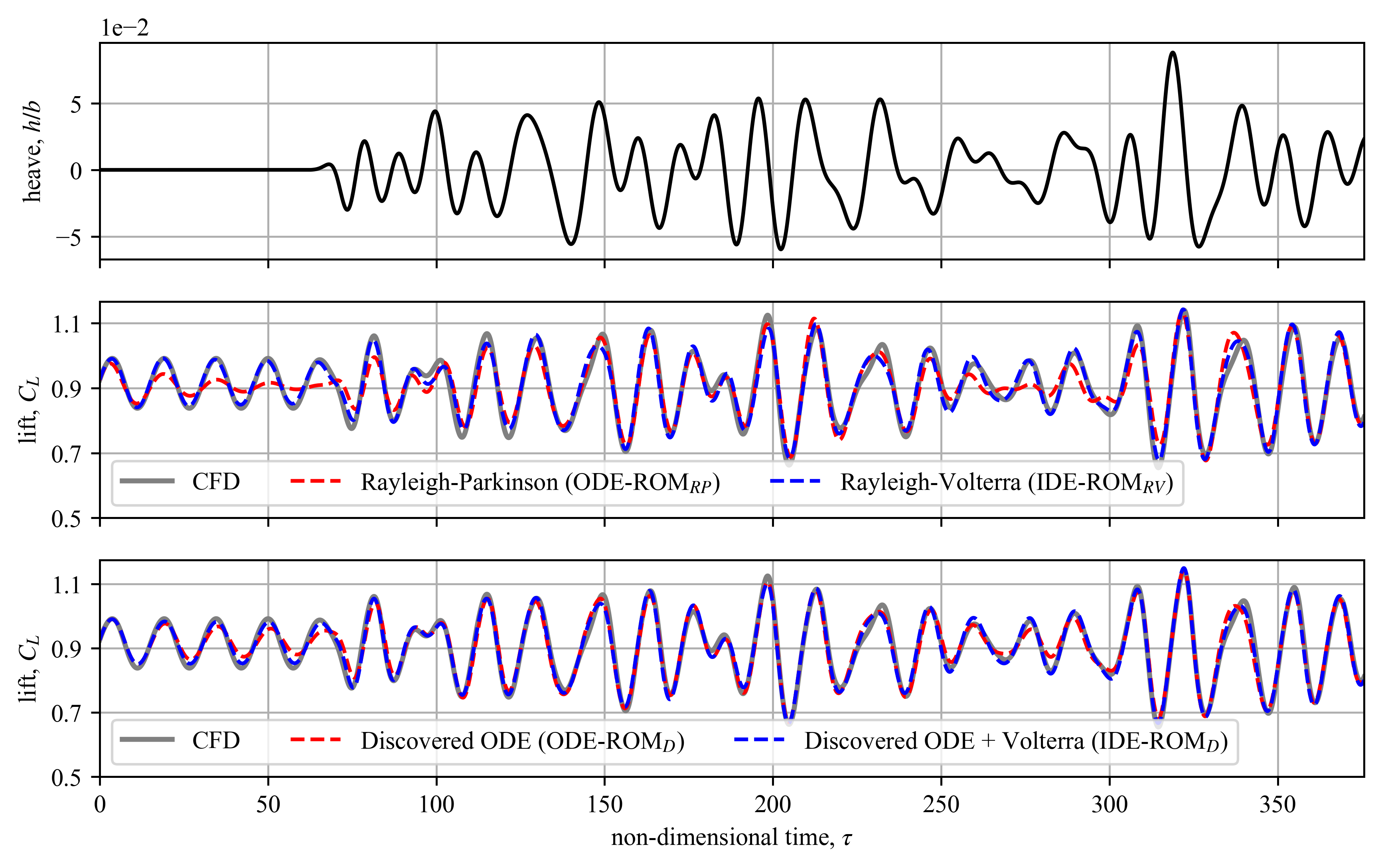}
		\caption{Time-integrated cross-validation data for lift due to heave motion (40k samples).}
		\label{fig:crossval_heave_lift}
	\end{figure}

\begin{figure}[h!]
		\centering
			\includegraphics[width=1\textwidth]{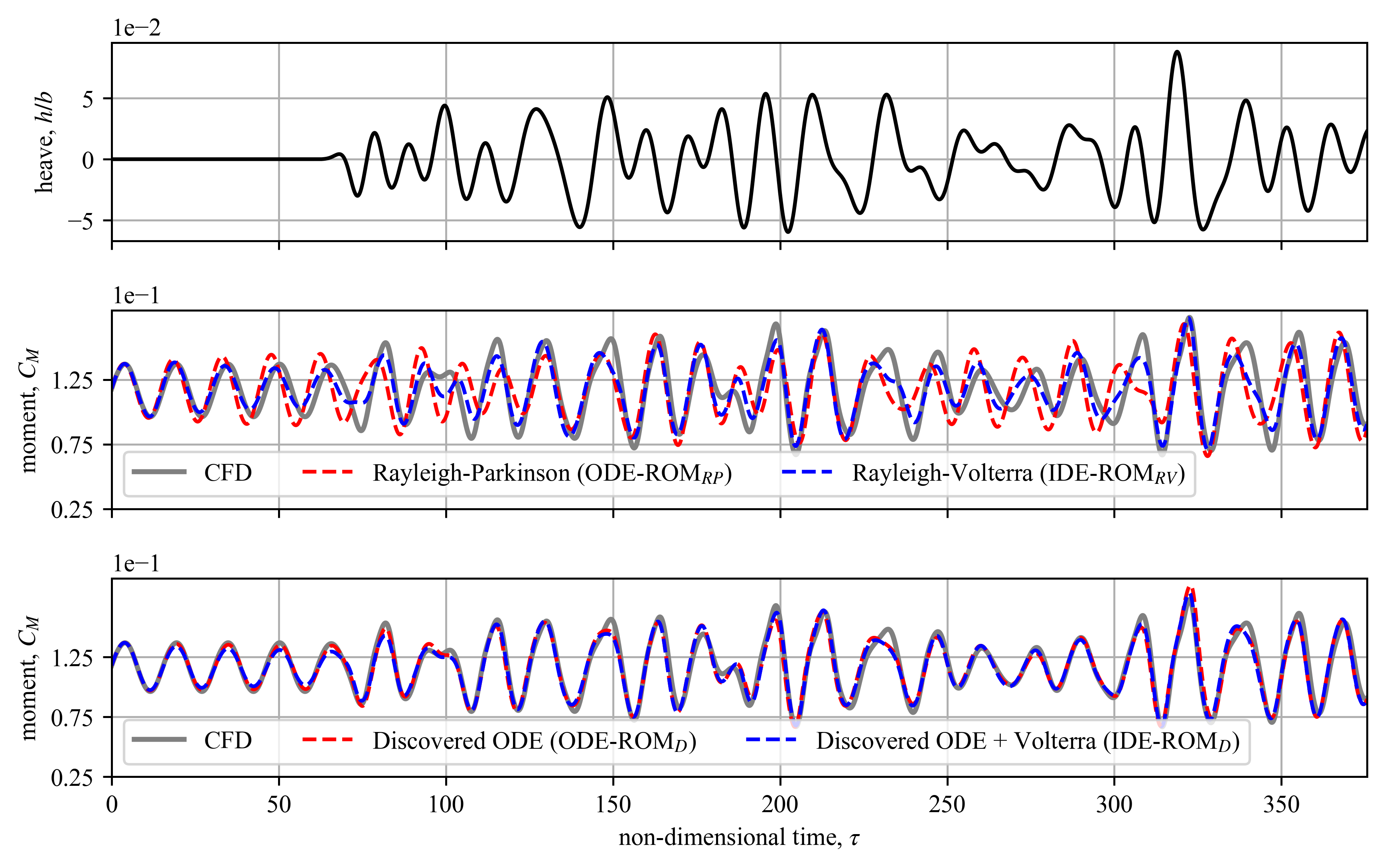}
		\caption{Time-integrated cross-validation data for moment due to heave motion (40k samples).}
		\label{fig:crossval_heave_mom}
	\end{figure}

    Training of the fixed-equation ODE ROMs ($\text{\textbf{ODE-ROM}}_{RP}$) does not require any grid search. For the discovered ODE ROM ($\text{\textbf{ODE-ROM}}_{D}$) a grid search is conducted within the range  $5 \leq \kappa^* \leq 50$ to identify the optimal ODE (with $\kappa^*$ terms). For the IDE ROMs a grid search is conducted of the number of time lags in $\dot{h}$  $100 \leq N_L \leq 1200$ and the total number of coefficients $\kappa^* \leq \kappa_{TOT} \leq 100$. The identified ROM statistics and cross-validation normalized root mean square deviations (nrmsd) after time integration are presented in Table~\ref{tab:crossval_heave}. The first 40k samples of the time-integrated cross-validation predictions are presented in Fig.~\ref{fig:crossval_heave_lift} for the lift predictions and Fig.~\ref{fig:crossval_heave_mom} for the moment predictions. 
    
    Starting with the lift forces, it is encouraging that the identified Rayleigh-Parkinson ROM ($\text{\textbf{ODE-ROM}}_{RP}$) alone is able to predict the nonlinear fluid-structure interactions with reasonable accuracy ($nrmsd = 6.34\%$). The problem is that the fluid-only LCO is not captured, as can be seen in Fig.~\ref{fig:crossval_heave_lift}, where the fluid behaves as a damped harmonic oscillator. This means that the coefficients of the Rayleigh oscillator are not identified with correct signs ($i.e.$, no negative linear damping). The impact of swapping the Parkinson Galloping model with a fifth-order pruned Volterra series ($\text{\textbf{IDE-ROM}}_{RV}$) is significant, providing a 26\% decrease in cross-validation error, and allowing the fluid-only LCO to be captured well by the Rayleigh oscillator terms. It seems that by including time lags for the fluid-structure interactions it alleviates the burden placed on the fluid ODE in fitting the global system. The discovered ODE ROM ($\text{\textbf{ODE-ROM}}_{D}$) has 30 terms, including a fluid oscillator and those that describe the fluid-structure interactions, providing a 40\% decrease in cross-validation error compared to the Rayleigh-Parkinson ROM ($\text{\textbf{ODE-ROM}}_{RP}$) and a significantly improved prediction of the fluid-only LCO. Finally the discovered IDE ROM ($\text{\textbf{IDE-ROM}}_{D}$) which adds a pruned Volterra series to $\text{\textbf{ODE-ROM}}_{D}$, while also allowing the ODE coefficients to be re-computed, performs with high accuracy, yielding $nrmsd = 2.45\%$ and a well captured fluid oscillator.

    Next looking at the moments, the trends in terms of accuracy are largely similar, although the errors are approximately double those of the lift force. This is to be expected given that the pitching moment (taken about the quarter-chord location) is more sensitive to shock oscillations, and as a result nonlinearities are more pronounced in the pitching moment time-series. In this case, the inclusion of the Volterra series is even more important. For instance, the Rayleigh-Parkinson ROM performance is clearly unacceptable with $nrmsd>15\%$, and the error is reduced by 46\% through the addition of the pruned Volterra series in $\text{\textbf{IDE-ROM}}_{RV}$. Similar to the case of lift, the discovered ODE ROM  ($\text{\textbf{ODE-ROM}}_{D}$) is the second best performer. The discovered IDE ROM ($\text{\textbf{IDE-ROM}}_{D}$) reduces $nrmsd<5\%$ which may be acceptable, although the aeroelastic cases are the real test of ROM performance. Qualitative observation of the time-series for the discovered ROMs in Fig.~\ref{fig:crossval_heave_mom} demonstrates reasonably good performance including a well-captured shock buffet-only oscillations.

    \begin{table}[h]
    \caption{Identified ROM statistics and cross-validation error for aerodynamic forces due to heave motion.}
    \label{tab:crossval_heave}
    \begin{tabular}{ccccccccc}
    \hline
     & \multicolumn{4}{c}{Lift} & \multicolumn{4}{c}{Moment} \\
        \hline
         & $\kappa^*$ & $N_L$ & $\kappa$ & NRMSD [\%] & $\kappa^*$ & $N_L$ & $\kappa$ & NRMSD [\%] \\
        \hline
        $\text{\textbf{ODE-ROM}}_{RP}$ & 9& - & 9 & 6.34  & 9 & - & 9 & 15.69  \\
        $\text{\textbf{IDE-ROM}}_{RV}$ & 5 & 900 & 36 & 4.64 & 5 & 1000 & 10 & 8.54\\
        $\text{\textbf{ODE-ROM}}_{D}$ & 30& - & 30 & 3.82 & 26& - & 26 & 6.03  \\
        $\text{\textbf{IDE-ROM}}_{D}$ & 30& 1200 & 49 & 2.45 & 26& 300 & 39 & 4.83 \\
        \hline
    \end{tabular}

    \end{table}

    \clearpage

    \subsubsection{Aeroelastic Response}
    \label{sec:ae_response_heave}

\begin{figure}[h!]
        \centering
            \includegraphics[width=1\textwidth]{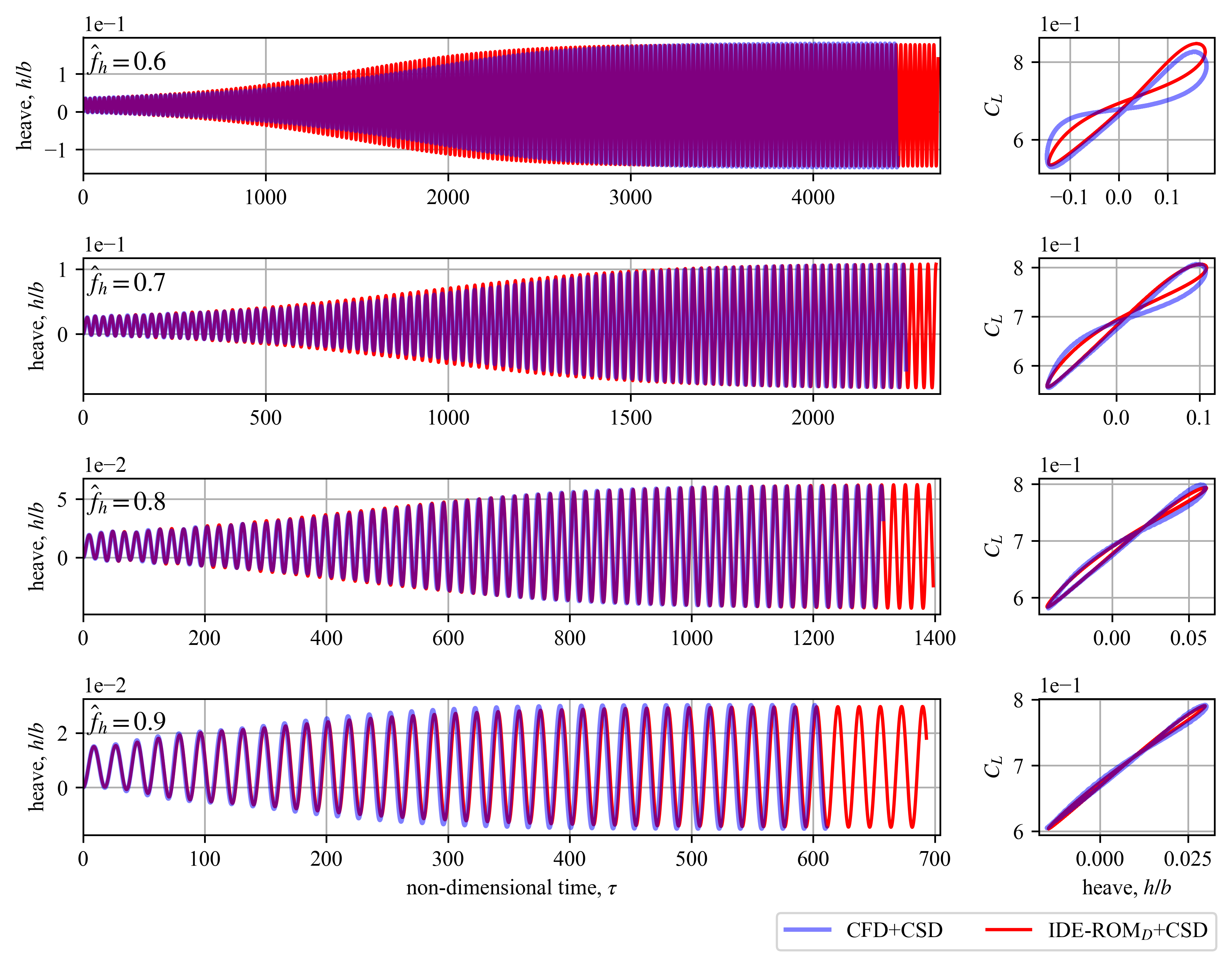}
        \caption{S-DOF heave aeroelastic responses with $\zeta_h = 0.000$ and various natural frequencies.}
        \label{fig:heave_time}
    \end{figure}

    The aeroelastic responses are now computed by coupling the discovered IDE ROM with the s-DOF heave structural equation of motion and marching the system forward in time. CFD-based aeroelastic simulations are also performed for verification. Initially the conventional analysis is conducted whereby the structural natural frequency is varied in order to map the region for which lock-in occurs as is presented in Fig.~\ref{fig:heave_time}. The system is modeled with structural damping ratios of $\zeta_h= 0.00, \, 0.005, \, 0.010$. The heave lock-in region, which may also be considered as a flutter due to the coupling of the structural and aerodynamic modes, commences near a frequency ratio of one and extends for $\hat{f}<1$, exactly the opposite to the curve that has been so often described for s-DOF pitch motion. Without structural damping, for the range of values tested, lock-off does not occur. The ROM predictions are excellent relative to the CFD/CSD result predicting the amplitude and frequency of the LCO with high accuracy. The most impressive aspect of this result is the ability of the ROM to extrapolate well beyond the training amplitude. A selection of time-series and Lissajous curves for this system are presented in Fig.\ref{fig:heave_time}. Again, the predictions are most encouraging; where not only are the amplitudes well predicted but also the transient component of the response (growth rate of the oscillations towards the stable limit cycle). The influence of structural damping is significant, as is to be expected. The ROM captures this large influence reasonably well, noting that the exact point of lock-off is challenging to get right given the fine balance between aerodynamic and structural damping. With $\zeta_h = 0.005$, lock-off is predicted by the CFD model to occur at $\hat{f}_h = 0.74$ while the ROM underpredicts by 6.8\% at $\hat{f}_h = 0.69$. With $\zeta_h = 0.010$ the lock-in region is small and does not extend beyond the intermittent region. The CFD-based prediction of lock-off here is $\hat{f}_h = 0.86$ while the ROM underpredicts this by 4.7\% at $\hat{f}_h = 0.82$.

    \clearpage
    
    \begin{figure}[h]
        \centering
            \includegraphics[width=1\textwidth]{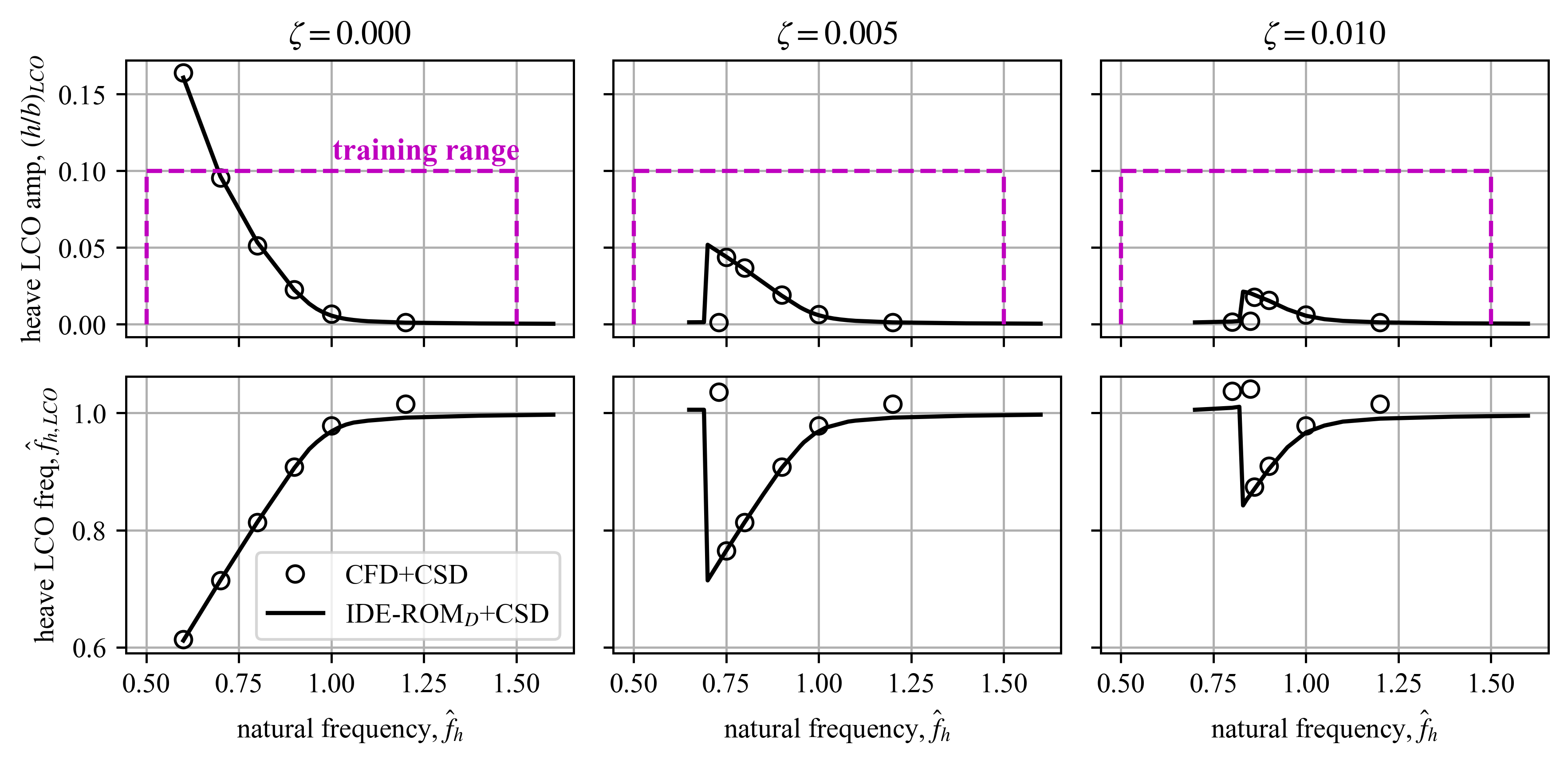}
        \caption{LCO amplitude and frequency for a s-DOF heave natural frequency sweep with different levels of structural damping. }
        \label{fig:heave_freq_sweep}
    \end{figure}

    \subsection{Lock-In/Flutter Onset}
        Contrary to conventional heave-pitch flutter where an increase of dynamic pressure triggers flutter due to a change in effective stiffness and coalescence of modes, here the driving mechanism is best thought of in terms of the structural-to-fluid mass ratio and the total effective damping (including structural and aerodynamic contributions). Instability comes not from stiffness loss but from synchronization (phase locking) between the structural mode and the fluid mode, occurring when the total effective damping becomes negative. In terms of the ROM predictions, a change in dynamic pressure (through the fluid density), or a proportional change to the mass of the wing, both have the same influence on the prediction of lock-in. However, for CFD-based predictions (or in experiment), this is not necessarily the case as the buffet frequency and amplitude vary based on Reynolds number -- making this test of the ROMs capacity particularly interesting. 
        
\begin{figure}[h]
        \centering
            \includegraphics[width=0.75\textwidth]{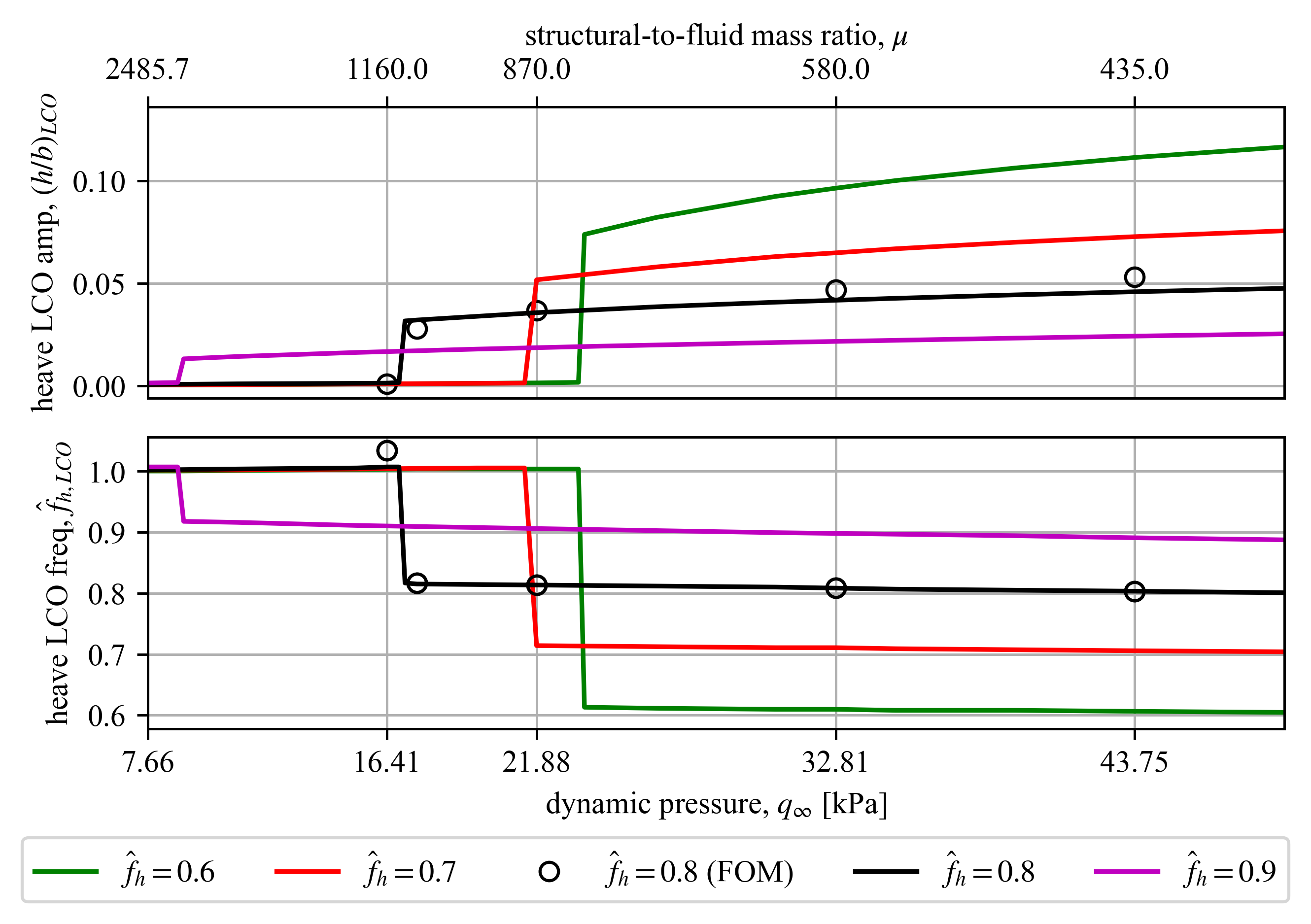}
        \caption{Dynamic pressure sweep for the S-DOF heave aeroelastic response with $\zeta_h = 0.005$ and various natural frequencies.}
        \label{fig:heave_flutter}
    \end{figure}

        Figure~\ref{fig:heave_flutter} presents the LCO amplitude and frequency as a function of dynamic pressure and mass ratio (which vary by fluid density). It can be seen that flutter/lock-in appears as a subcritical instability where the frequency abruptly shifts from the buffet frequency to the structural natural frequency. As the structural natural frequency reduces, higher dynamic pressures (lower mass ratios) are required to trigger flutter/lock-in. However, when it does occur, it is of a higher-amplitude. Comparison of the CFD-based and ROM-based computations at ($\hat{f}_h = 0.8$) demonstrate a surprisingly good prediction of the flutter/lock-in dynamic pressure and the amplitude of the LCO for the dynamic pressure sweep. This is despite the frequency and the magnitude of the buffet response changing in the CFD model but not the ROM. For instance, at the training conditions ($q_\infty = 21.88$kPa, $Re_\infty = 3\times10^6$) the buffet frequency is $f_B = 74.2$Hz and $\Delta C_L = 0.15$, while at $q_\infty = 43.75$kPa, $Re_\infty = 6\times10^6$ the buffet frequency is $f_B = 79.3$Hz and $\Delta C_L = 0.19$. 
        
        Another interesting note is that the flutter mass ratio appears to grow exponentially as a function of the structural natural frequency. This can be interpreted as the required airfoil mass to suppress lock-in going to infinity (or the fluid density going to zero) as the structural natural frequency approaches the buffet frequency. This trend was described by Gao $\textit{et al.}$~\cite{gao20} who discuss the "boot-like shape" of the aeroelastic stability region and will be explored further next.  

    \subsubsection{Critical Mass Ratio}
    \label{sec:crit_mu}
\begin{figure}[h!]
        \centering
            \includegraphics[width=1\textwidth]{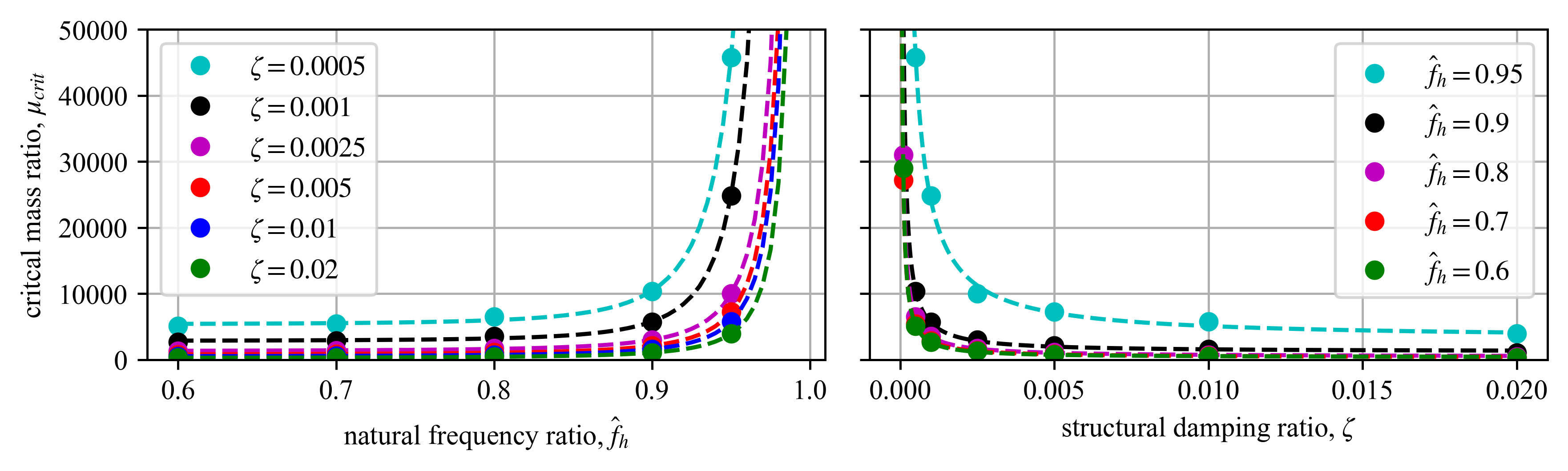}
        \caption{Critical mass ratio with various structural damping ratios and natural frequencies}
        \label{fig:heave_crit_mass}
    \end{figure}

    Now the ROM-based aeroelastic model is used to further investigate the critical mass ratio, $\mu^*$, defined here as the mass ratio required to suppress lock-in/flutter. Figure~\ref{fig:heave_crit_mass} presents the critical mass ratio as a function of the structural natural frequency and damping ratio, where two clear behaviors can be observed: 
     \begin{itemize}
         \item \textbf{Observation 1}: The mass ratio required to suppress lock-in / flutter tends to infinity as the structural natural frequency ratio approaches one.
         \item \textbf{Observation 2}: The mass ratio required to suppress lock-in / flutter tends to infinity as the structural damping ratio approaches zero.
     \end{itemize} 
     
    An explanation of \textbf{Observation 1} is intuitive when viewing the structural mode as a weakly forced oscillator driven by the buffet limit cycle. As the structural natural frequency approaches the buffet frequency ($\omega_h/\omega_B \rightarrow 1$), the detuning $\Delta\omega = \omega_B - \omega_h$ tends to zero. In this regime, the relative phase between buffet and structural motion,  $\phi = \phi_B - \phi_h$, can be described by the classical Adler equation~\cite{adler46} for a forced self–excited oscillator: 
    
    \begin{equation}
    \dot{\phi} = \Delta\omega - K \sin\phi
    \end{equation}
    
    \noindent where $K$ is an effective coupling gain measuring how strongly the buffet oscillator drives the structural mode. The Adler condition for phase locking is $|K| \gtrsim |\Delta\omega|$: when the coupling exceeds the detuning, the phase $\phi$ becomes trapped and the two oscillators synchronize. To avoid synchronization it is therefore required that $|K| \lesssim |\Delta\omega|$. 
    
    In the present setting, the coupling gain scales as $K \propto L_0 |H(\omega_B)|$, where $L_0$ is the buffet lift amplitude and $H(\omega)$ is the structural frequency–response function. Since the FRF magnitude $|H(\omega_B)|$ decreases with increasing effective mass, the only way to keep $|K|$ below the shrinking detuning $|\Delta\omega|$ as $\omega_h \rightarrow \omega_B$ is to increase the effective mass, and thus the mass ratio $\mu$. In other words, as the structural natural frequency approaches the buffet frequency, the system becomes increasingly prone to phase locking, and an ever larger mass ratio is required to desensitize it to buffet–induced synchronization. Observation 1 therefore reflects a resonance–driven amplification of the coupling between buffet and structural motion, such that, in the limit $\omega_h/\omega_B \rightarrow 1$, the mass ratio needed to suppress lock–in diverges as
    
    \begin{equation}
        \label{eq:res11}
        \mu \gtrsim \frac{\text{const}}{|\Delta\omega|} \;\;\rightarrow\;\; \infty 
        \quad \text{as} \quad \omega_h/\omega_B \rightarrow 1 
    \end{equation}

    
    
    
    
    

    Now, in terms of \textbf{Observation 2}, the lock-in/flutter mechanism is investigated from the perspective of the aerodynamic damping or, more specifically, the total effective damping $c_\text{eff}$, which can be negative or positive. Negative effective damping can occur when the aerodynamic damping, $c_{a_h}$, is positive (in the convention used herein) and larger than the structural damping, $c_h = 2\zeta_h\omega_h m$. To quantify this, the unsteady aerodynamic forces are recorded under forced harmonic excitation, then the work–per–cycle (WPC) method is used to compute the aerodynamic damping. The aerodynamic damping coefficient is obtained from the net work of the aerodynamic force~\cite{vasanthakumar11}, $L_{H}(t)$, over an integer number of forced harmonic excitation cycles, $h_{H}(t)$, at frequency $\omega$:
    
    \begin{equation}
      W = \int_0^T L_{H}(t)\,\dot h_{H}(t)\, dt
    \end{equation}
    
    \begin{equation}
      c_{a_h} = \frac{W}{\pi \hat{h}^2 \omega}
    \end{equation}
    
    \noindent where $\hat{h}$ is the amplitude of $h_{H}(t)$. In this convention, $c_{a_h}>0$ is destabilizing (energy input to the structure), while $c_{a_h}<0$ indicates dissipative behavior.

\begin{figure}[h]
        \centering
            \includegraphics[width=1\textwidth]{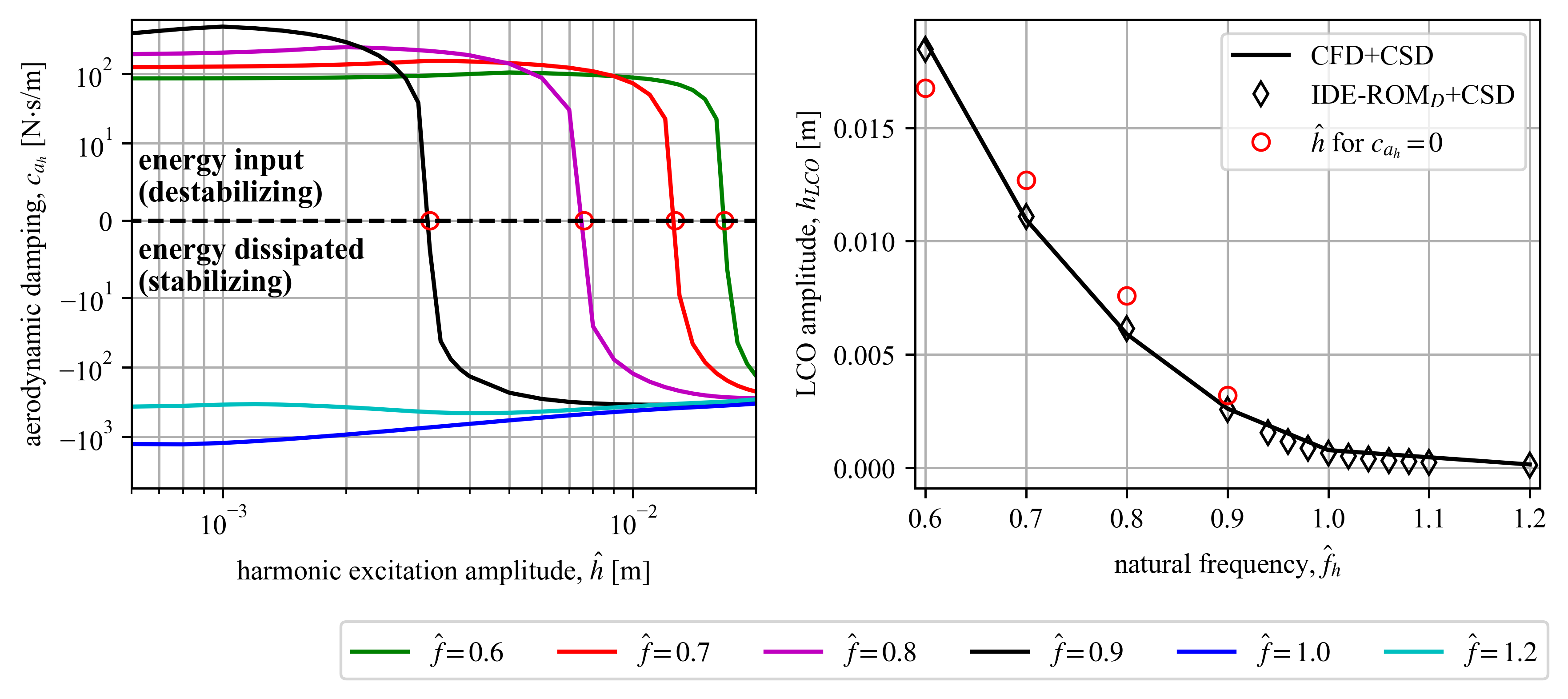}
        \caption{Aerodynamic damping estimates for various heave harmonic excitation frequencies and amplitudes.}
        \label{fig:heave_damping}
    \end{figure}

    Figure~\ref{fig:heave_damping} presents the aerodynamic damping as a function of the amplitude of the forced harmonic excitation in heave for a range of excitation frequencies. It is clear that for excitation frequencies $\hat{f} < 1$, when the amplitude of the structural vibrations are small, aerodynamic damping values are consistently positive. This means that when the aerodynamic damping is greater than the structural damping, the effective damping is negative ($c_\text{eff} = c_h - c_{a_h} < 0 \, \forall \, c_{a_h} > c_h$). This condition is guaranteed when the structural damping is zero $c_h = 0 \implies \forall c_{a_h} > 0: \, c_\text{eff} < 0$. This explains the explosion of the critical mass ratio as structural damping approaches zero: there is no mass that can suppress lock-in when the effective damping is negative. Moreover, the aerodynamic damping seems to grow exponentially as $\hat{f}\rightarrow 1$, suggesting that for the aeroelastic model the structural damping required to suppress lock-in becomes very large as $\hat{f}_h\rightarrow 1$ -- aligned with the aeroelastic behaviors observed in Fig.~\ref{fig:heave_freq_sweep}. Also of interest are the aerodynamic damping values for $\hat{f} > 1$ where it can be seen that the aerodynamic damping is negative, guaranteeing positive effective damping ($c_\text{eff} = c_h - c_{a_h} > 0, \, \forall \, c_{a_h} < 0$). Again, this aligns with what is observed in Fig.~\ref{fig:heave_freq_sweep} where lock-in does not occur for structural natural frequencies $\hat{f}_h > 1$. 
    
    Another interesting property of Fig.~\ref{fig:heave_damping} is the zero crossing that occurs with larger amplitude excitation, suggesting that at some point, when the structural oscillation amplitude is large enough, the aerodynamics has a stabilizing effect (dissipates vibrational energy from the structure). This aligns with what is well known to occur in aeroelastic systems, where a balancing of the aerodynamic and structural forces result in LCO. Comparison of the values of $\hat{h}$ at zero aerodynamic damping to the aeroelastic LCO amplitudes are well correlated, indicating that computation of aerodynamic damping may be a viable approach to not only investigate stability, but also to approximate the aeroelastic LCO amplitude.

    Overall, these results comprehensively show that in the case of shock buffet, the lock-in aeroelastic instability is driven by negative effective damping and is perhaps better thought of as a s-DOF flutter. This not only explains the high sensitivity of the aeroelastic system to even small amounts of structural damping, but also explains the sensitivity to mass ratio. Specifically, when the structural mass is large enough (or fluid density small enough) such that $c_h > c_{a_h}$, the lock-in/flutter instability can be completely suppressed. Otherwise the increase in mass ratio affects only the number of cycles for a stable LCO to develop, while the properties of the LCO itself are largely unaffected. These behaviors are similar to those described for a s-DOF transonic flutter mechanism~\cite{dowell24}.

    \subsection{Aeroelastic Reduced-Order Model Differential Equations: Pitch Motion}

    This section is concerned with modeling the s-DOF pitch response of the wing under shock buffet excitation. Only full-order simulations and discovered ROMs are considered. As per the training description in Section~\ref{sec:heave_train}, an initial grid search is conducted within the range  $5 \leq \kappa^* \leq 50$ to discover the base ODE ($\text{\textbf{ODE-ROM}}_D$). Then the IDE ROM is identified ($\text{\textbf{IDE-ROM}}_D$) by adding time lags in $\dot{\alpha}$ and conducting a grid search of the number of time lags $100 \leq N_L \leq 1200$ and the total number of coefficients $\kappa^* \leq \kappa_{TOT} \leq 100$ (where $\kappa^*$ is the number of terms in the base ODE selected from the original grid search). The addition of lags in $\alpha$ rather than $\dot{\alpha}$ was also tested given that it is a conventional formulation for nonlinear unsteady aerodynamic ROM in the pitching mode~\cite{balajewicz12}. However, the $\dot{\alpha}$-based formulation is found to provide a slightly more accurate model given that the unsteady aerodynamic forces during shock buffet on an oscillating airfoil are more sensitive to motion dynamics than to the instantaneous pitch angle.

\begin{figure}[h]
		\centering
			\includegraphics[width=1\textwidth]{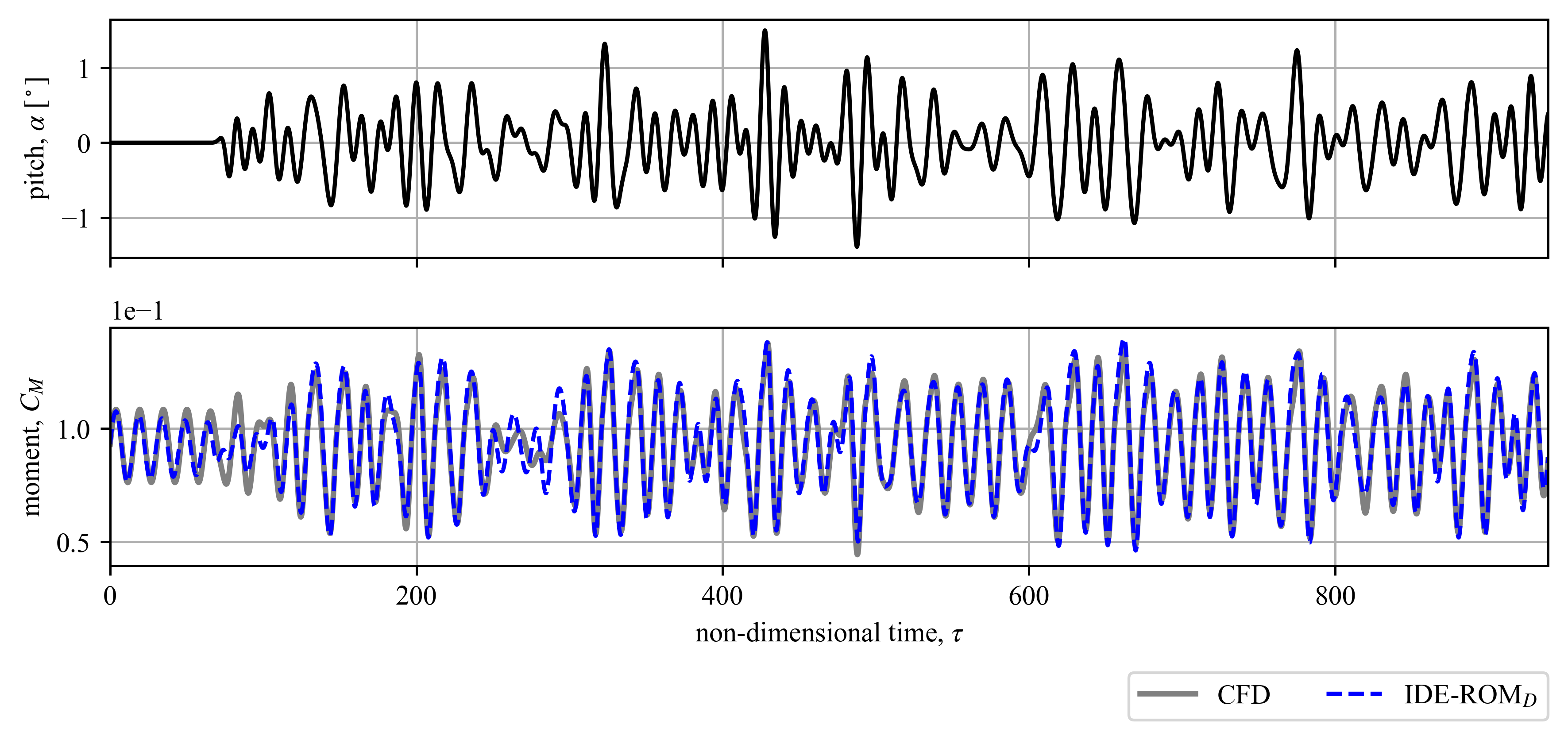}
		\caption{Time-integrated cross-validation data for moment due to pitch motion (100k samples).}
		\label{fig:crossval_pitch_mom}
	\end{figure}

    The cross-validation error of the $\text{\textbf{ODE-ROM}}_D$ is $nrmsd = 6.02$\% and the $\text{\textbf{IDE-ROM}}_D$ is $nrmsd = 4.85$\%. The first 100k samples of the time-integrated $\text{\textbf{IDE-ROM}}_D$ cross-validation predictions are presented in Fig.~\ref{fig:crossval_pitch_mom}. It is clear that the high amplitude moment fluctuations are well captured, while the fluid (buffet) only oscillations and some regions of low structural excitation (where pure buffet dynamics dominate) are erroneous. It can be seen that in the initial region without structural dynamics, the predicted oscillations are decaying. Although not shown, the fluid-only component of the discovered oscillator does not decay to a stable response, but rather the amplitude of the predicted buffet cycle is approximately half that of the true value. As discussed in Section~\ref{sec:heave_train}, this significant increase in error compared to the modeling of lift forces is expected. The major concern surrounds the accuracy of aeroelastic predictions which will be assessed next.

        \clearpage

\begin{figure}[h]
        \centering
            \includegraphics[width=1\textwidth]{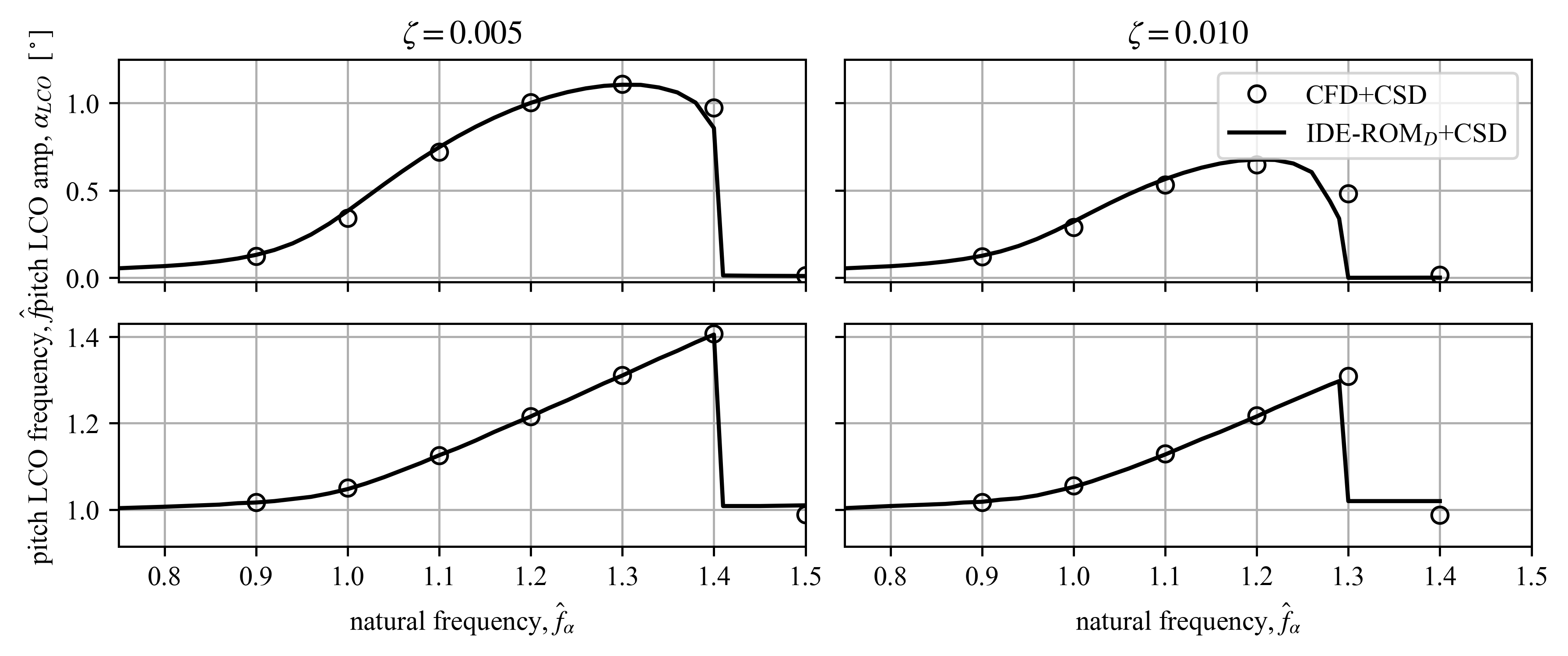}
        \caption{LCO amplitude and frequency for a S-DOF pitch natural frequency sweep with different levels of structural damping. }
        \label{fig:pitch_freq_sweepA}
    \end{figure}

\begin{figure}[h]
        \centering
            \includegraphics[width=1\textwidth]{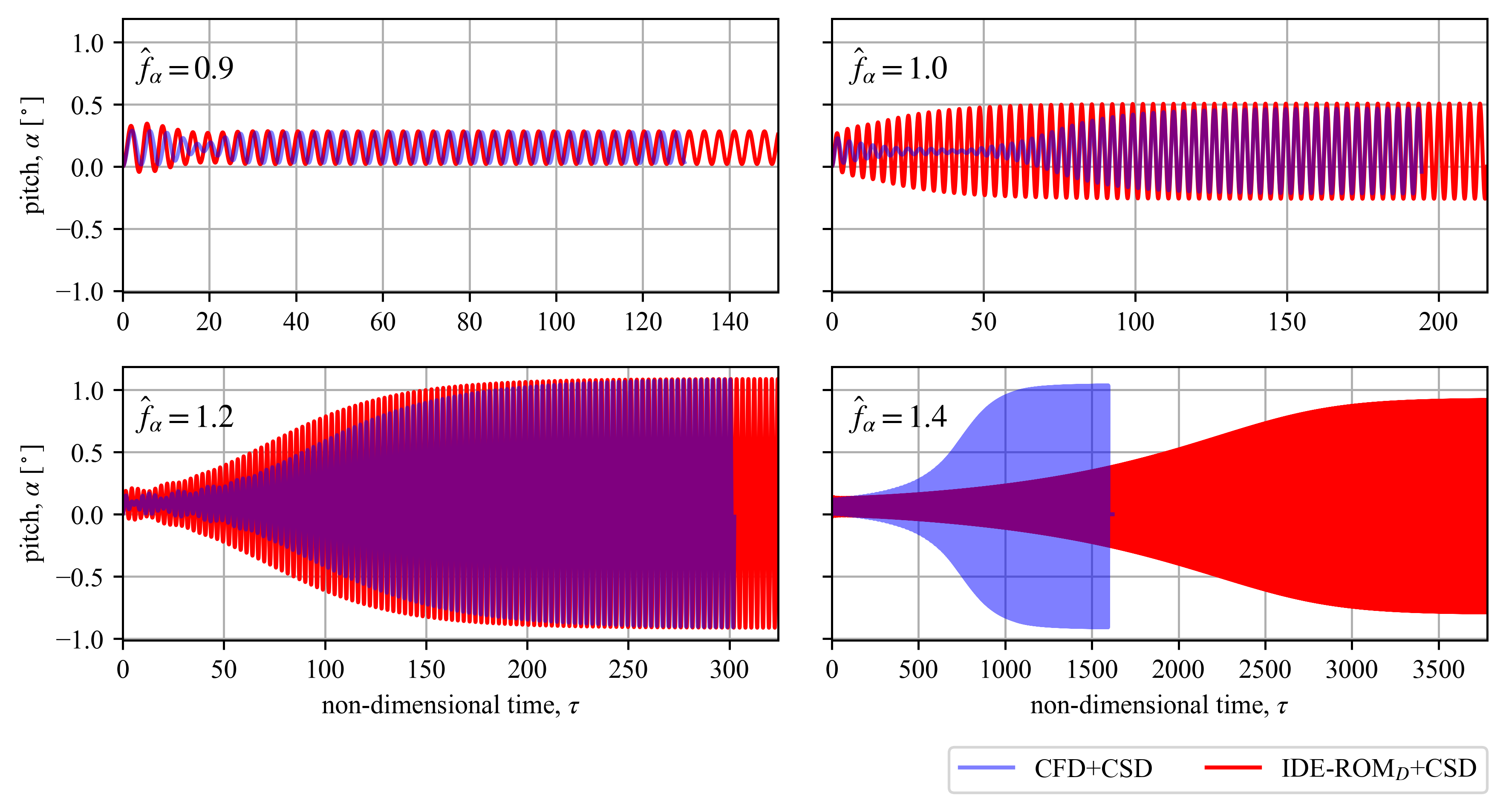}
        \caption{S-DOF pitch aeroelastic responses with $\zeta_\alpha = 0.005$ and various natural frequencies.}
        \label{fig:pitch_time}
    \end{figure}

    Figure~\ref{fig:pitch_freq_sweepA} presents the s-DOF pitch aeroelastic LCO amplitudes and frequencies for a sweep of the structural natural frequencies. Solutions are computed by coupling the discovered IDE ROM with the s-DOF pitch structural equation of motion and marching the aeroelastic system forward in time. CFD-based aeroelastic simulations are performed for verification. By varying the structural natural frequency, the region for which lock-in occurs is mapped for structural damping ratios of $\zeta_\alpha = 0.005, \, 0.010$. These structural damping values lead to aeroelastic responses that do not extend outside the range of the training data. As has also been shown by many other authors~\cite{gao20}, lock-in is triggered at $\hat{f}_\alpha \approx 1$ and extends to some value $1<\hat{f}_\alpha<2$. This is exactly the opposite to what is observed for s-DOF heave motion in the previous Section~\ref{sec:ae_response_heave} (related to aerodynamic damping as will be demonstrated later in this section). It is clear that the ROM predicts the lock-in LCO amplitude and frequency with reasonably high accuracy, despite the reduced accuracy aerodynamic model (in comparison to the model for heave in the previous Section~\ref{sec:aeroelastic_heave}), which is certainly encouraging. The ROM prediction of lock-off for $\zeta_\alpha = 0.005$ is $\hat{f}_\alpha = 1.41$ (2.76\% less than the true value), and for $\zeta_\alpha = 0.010$ is at $\hat{f}_\alpha = 1.29$ (5.15\% less than the true value). The time responses are presented in Fig.~\ref{fig:pitch_time} where the influence of the diminished accuracy of the aerodynamic model is clearer. Notably, the transients are poorly captured and, as the natural frequency approaches lock-off ($\hat{f}_\alpha \approx 1.4$), error in the growth rate becomes more pronounced. 
    
    \clearpage

\begin{figure}[h]
        \centering
            \includegraphics[width=1\textwidth]{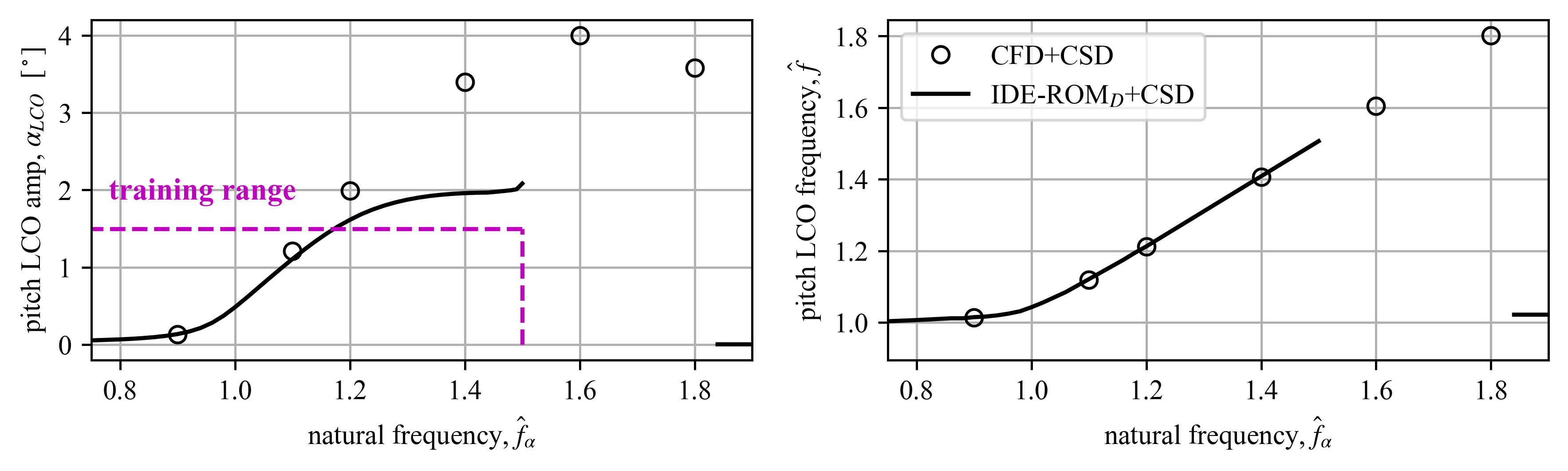}
        \caption{LCO amplitude and frequency for a S-DOF pitch natural frequency sweep with zero structural damping. }
        \label{fig:pitch_freq_sweepB}
    \end{figure}

    Next, the structural damping is set to zero and the ability of the ROM to extrapolate outside the amplitudes and frequencies observed in the training set are assessed. As can be observed in Fig.~\ref{fig:pitch_freq_sweepB}, in terms of predicting the LCO amplitude, the extrapolative performance is poor. At $\hat{f}_\alpha = 1.4$, the LCO amplitude is underpredicted by $\sim$70\%. Unsurprisingly, the LCO frequencies are captured well despite poor prediction of the amplitude. When the natural frequency extends beyond the maximum training frequency ($\hat{f}_\alpha = 1.5$), the ROM cannot predict LCO as the system becomes unstable. This is reflected in the gap from $1.5 < \hat{f}_\alpha < 1.84$. At $\hat{f}_\alpha = 1.85$ the ROM predicts lock-off which is $\sim 2\%$ greater than ground truth. Overall these findings are not unexpected, highlighting that for the accurate prediction of the LCO amplitude, while some extrapolation may be possible, the training data should be designed to capture the range of amplitudes and frequencies seen in the aeroelastic system.
    \clearpage

\begin{figure}[h]
        \centering
            \includegraphics[width=0.75\textwidth]{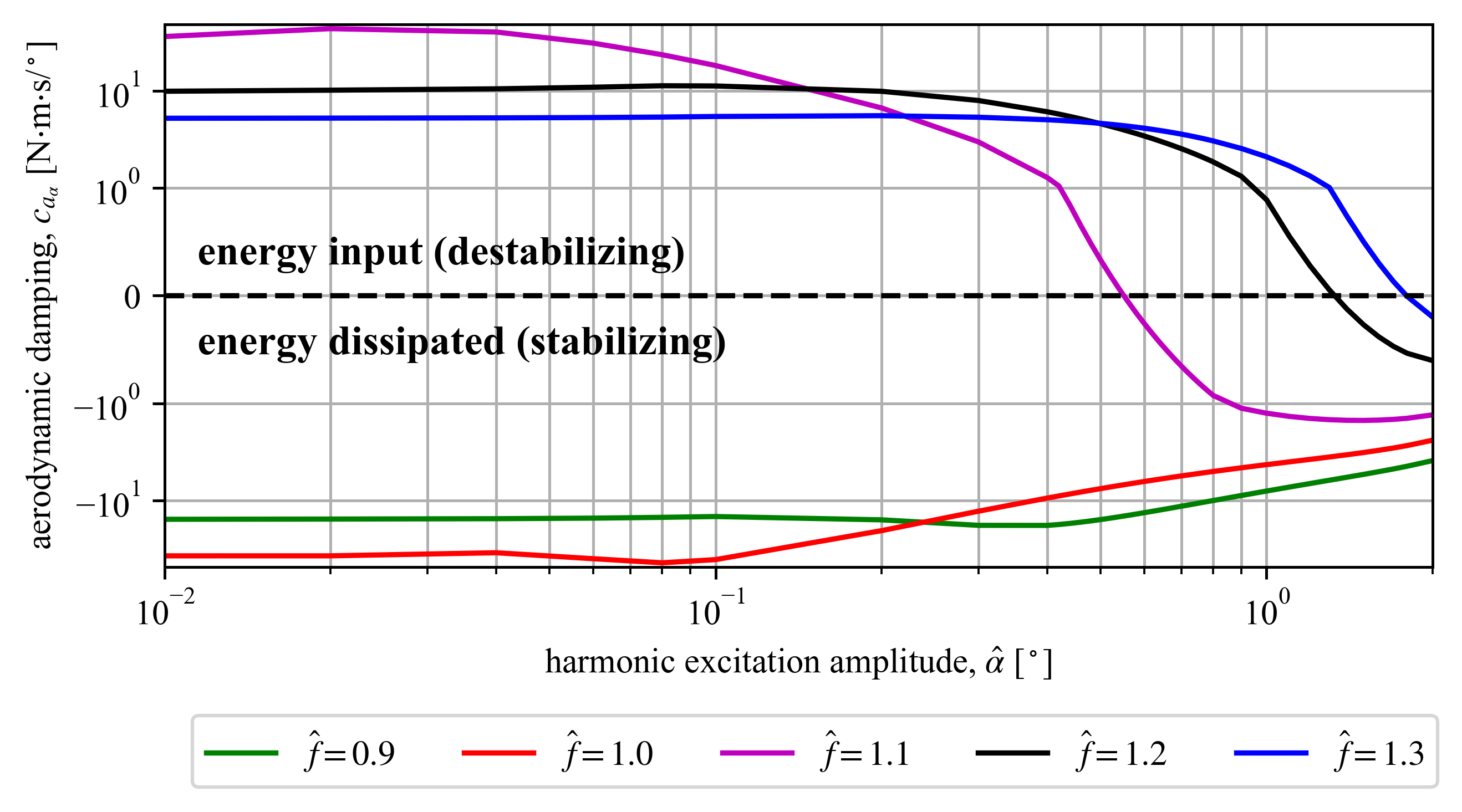}
        \caption{Aerodynamic damping estimates for various pitch harmonic excitation frequencies and amplitudes. }
        \label{fig:pitch_damping}
    \end{figure}

    Finally, the aerodynamic damping is assessed for the s-DOF pitching system as presented in Fig.~\ref{fig:pitch_damping}. The methodology used is the same as Section~\ref{sec:crit_mu}. The harmonic pitching amplitudes, $\hat{\alpha}$, range from $0.01^\circ \leq \hat{\alpha} \leq 2^\circ$. Much like the aeroelastic lock-in trends, the aerodynamic damping trends are exactly reversed compared to the s-DOF heave case. Specifically, it can be seen that for low harmonic excitation amplitudes, when $\hat{f}>1$ the aerodynamic damping is positive, and when $\hat{f}<1$ the aerodynamic damping is negative. This means that negative effective damping ($c_{eff}  = 2\zeta_\alpha\omega_\alpha I_\alpha - c_{a_\alpha} < 0$) only occurs for $\hat{f}>1$, which suggests that lock-in is only possible for natural frequencies $\hat{f}_\alpha>1$ (confirmed in Figs.~\ref{fig:pitch_freq_sweepA} and~\ref{fig:pitch_freq_sweepB}). The harmonic excitation amplitude at zero damping is not as well correlated with the LCO amplitude as was observed for heave motion in Fig.~\ref{fig:heave_damping}, underpredicting by a factor of 2. This may be the result of the reduced accuracy aerodynamic model and requires further investigation. Overall, these results strongly support the observation that lock-in is driven by negative effective damping. The reader is referred to the previous Section~\ref{sec:crit_mu} for a more comprehensive discussion on this mechanism.

\subsection{Computational Savings}

One of the primary motivations for developing the IDE-ROM framework is to reduce the prohibitively high cost of high-fidelity CFD/CSD aeroelastic simulations in transonic buffeting flow. Each fully coupled two-dimensional simulation typically requires 100,000--2,000,000 time steps to reach a stable limit cycle, consuming approximately 600–12,000 CPU-hours per simulation on 30 cores. In contrast, once trained, the IDE-ROM can be integrated forward in time using a simple explicit scheme on a single CPU core, requiring only a fraction of a CPU-hour. This corresponds to an approximate speed-up of 10,000$\times$--100,000$\times$ relative to the high-fidelity reference for the presented case studies. For the present study, the initial cost of generating the training data are approximately 1,000 CPU-hours, meaning that computational savings are observed after a single aeroelastic simulation. As mentioned previously, the number of training samples can almost certainly be substantially reduced, however, it is outside of the scope of this work. These computational savings make the proposed IDE-ROM approach particularly attractive for rapid stability mapping, uncertainty quantification, and integration within digital twin frameworks for transonic aeroelastic systems.

OMP provides an additional advantage for sparse model identification. In contrast to $\ell_1$-penalized methods such as LASSO, which rely on iterative optimization procedures to compute regularized solutions, OMP employs a greedy selection strategy with closed-form least-squares updates at each step. For problems with $\sim 10^{5}$ training samples and $\sim 10^{4}$-dimensional state libraries, OMP can substantially reduce training time while maintaining comparable sparsity and predictive accuracy under appropriate conditions (e.g., moderate noise levels and limited feature correlation). This computational efficiency is particularly beneficial when evaluating multiple model variants or performing grid searches over time-lag and sparsity parameters, thereby improving the overall efficiency of the framework.

\section{Conclusions}

This paper proposes and evaluates a nonlinear unsteady aerodynamic reduced-order modeling framework for transonic buffet aeroelasticity based on data-identified differential and integro-differential equations. The proposed nonlinear oscillator-with-memory combines a compact self-excited fluid oscillator with a pruned Volterra series representation of the nonlinear, history-dependent aerodynamic response to structural motion. Both prior-known and discovered formulations are presented, with governing terms and coefficients identified using orthogonal matching pursuit. Demonstrated on the ONERA OAT15A airfoil for buffet-only and aeroelastic cases, the method reproduces buffet limit cycles, captures lock-in regions, and predicts LCO amplitudes and frequencies in close agreement with CFD/CSD benchmarks. The lock-in instability is shown to be driven by negative effective damping, accounting for both aerodynamic and structural contributions. Further to this, it is shown that the aeroelastic stability of the system and the aeroelastic LCO amplitude can be extracted by computing the aerodynamic damping from simple harmonic excitation ($i.e.$, without running aeroelastic simulations). The primary limitations of the approach include:

\begin{enumerate}
  \item When the training input amplitudes exceed moderate levels (e.g., pitch motions $>2^\circ$), the cross-validation error increases sharply, indicating that the identified ROM no longer captures the nonlinear dynamics with sufficient fidelity.
  \item In fully three-dimensional buffet, the flow response becomes broadband and quasi-aperiodic, making it difficult to represent using compact oscillator-based formulations; additional modes or stochastic extensions are likely required.
\end{enumerate}

Overall, the IDE-ROM provides an interpretable and computationally efficient surrogate for transonic buffet aeroelasticity, and future work will extend it to multi-DOF, multi-input formulations, including embedding the identified governing equations within a Neural ODE framework to improve performance at high-amplitude excitation levels.

\section*{Acknowledgments}
This work would not have been possible without the financial support provided by the Asian Office of Aerospace Research and Development (AOARD) and Air Force Office of Scientific Research (AFOSR) for project FA2386-24-1-4044: Data-Driven Reduced Order modeling and Preliminary Experimentation for Combined Transonic Buffet and Freeplay Induced Limit Cycle Oscillations. We are grateful for the partial financial support provided by the Australian Defence Science and Technology Group (DSTG). The computational resources provided by ANSYS, and the support of Dr Valerio Viti and Dr Luke Munholand, are also greatly appreciated.
\bibliographystyle{elsarticle-num} 
\bibliography{gensys_doi}
        

\end{document}